\documentclass[useAMS,usenatbib]{mn2e}
\usepackage{graphicx}
\usepackage{graphicx,color}
\usepackage{textcomp}
\usepackage{amsmath}

\graphicspath{{Figures/}}
\DeclareGraphicsExtensions{.eps, .ps}

\renewcommand{\arcsec}{$^{\prime\prime}$}

\newcommand{\qm}[1]{``#1''}
\newcommand{\oiii}{[O\,{\scriptsize III}\,]}
\definecolor{gold}{rgb}{0.85,.55,0}
\definecolor{orange}{rgb}{1,0.5,0}

\title[Stellar population and gas kinematics of the PSQ J0210-0903]{2D stellar population and gas kinematics of the inner 1.5\,kpc
of the post-starburst quasar SDSS\,J0210-0903}

\author[David Sanmartim, Thaisa Storchi-Bergmann and Michael S. Brotherton]{David Sanmartim$^{1}$\thanks{E-mail:davidsanm@gmail.com}, 
Thaisa Storchi-Bergmann$^{1}$ and Michael S. Brotherton$^{2}$ \\
$^{1}$Universidade Federal do Rio Grande do Sul, IF, CP 15051, Porto Alegre 91501-970, RS, Brasil\\
$^{2}$University of Wyoming, Department of Physics and Astronomy, Laramie, WY, 82071, USA}
\begin{document}

\date{Accepted 2012 September 25.  Received 2012 August 24; in original form 2012 May 17}

\pagerange{\pageref{firstpage}--\pageref{lastpage}} \pubyear{2012}

\maketitle

\label{firstpage}

\begin{abstract}
Post-Starburst Quasars (PSQs) are hypothesized to represent a stage in the evolution of massive galaxies in which the star formation has been recently quenched due 
to the feedback of the nuclear activity. In this paper our goal is to test this scenario with a resolved stellar population study of the PSQ J0210-0903 ,as well as 
of its emitting gas kinematics and excitation. We have used optical Integral Field Spectroscopy obtained with the Gemini GMOS instrument at a velocity resolution of 
$\approx$\,120\,km\,s$^{-1}$ and spatial resolution of $\approx$\,0.5\,kpc. We find that old stars dominate the luminosity (at 4700\,\AA) in the inner 0.3\,kpc (radius), 
while beyond this region (at $\approx$\,0.8\,kpc) the stellar population is dominated by both intermediate age and young ionizing stars. The gas emission-line ratios 
are typical of Seyfert nuclei in the inner  0.3\,kpc, where an outflow is observed. Beyond this region the line ratios are typical of LINERs and may result from 
the combination of diluted radiation from the nucleus and ionization from young stars. The gas kinematics show a combination of rotation in the 
plane of the galaxy and outflows, observed with a maximum blueshift of $-670$ km s$^{-1}$. We have estimated a mass outflow rate in ionized gas in the range $0.3-1.1\, 
\rmn{M}_{\odot}$\,yr$^{-1}$ and a kinetic power for the outflow of $\dot{E}_\rmn{out} \approx 1.4-5.0 \times 10^{40}\,\rmn{erg}\,\rmn{s}^{-1} \approx 0.03\%-0.1\% \times
L_\rmn{bol}$. This outflow rate is two orders of magnitude higher than the nuclear accretion rate of $\approx\,8.7 \times 10^{-3}\,\rmn{M}_{\odot}$\,yr$^{-1}$, 
thus being the result of mass loading of the nuclear outflow by circumnuclear galactic gas. Our observations support an evolutionary scenario in which  the feeding 
of gas to the nuclear region has triggered a circumnuclear starburst 100's\,Myr ago, followed by the triggering of the nuclear activity, producing the observed gas 
outflow which may have quenched further star formation in the inner 0.3\,kpc.

\end{abstract}


\section{Introduction}
\label{sec:intro}

Our understanding of galaxy evolution has undergone a revolution in the past decade. Of particular interest is the correlation that links the nuclear supermassive black 
holes (SMBH) mass to the bulge stellar velocity dispersion, the $M_{\rm BH}-\sigma_*$ relation \citep{gebhardt00a,gebhardt00b,merritt01,tremaine02,ferrarese05,gultekin09,
graham11}. In this scenario, Post-Starburst Quasars (PSQs) may acquire a great importance, since they seem to represent a critical phase in the secular evolution of galaxies 
that links the growth of the stellar bulge and that of the SMBH.

PSQs are broad-lined AGN that show Balmer jumps and high-order Balmer absorption lines from A stars characteristic of massive post-starburst populations with age 
of a few hundred Myrs. PSQs are hypothesized to represent a stage in the evolution of massive galaxies in which both star formation and nuclear activity have been 
triggered and are visible simultaneously before one or the other fades. \cite{brotherton99} reported the discovery of a spectacular PSQ that they propose to represent
a stage in hierarchical galaxy evolution following a merger and enshrouded Ultra Luminous Infrared Galaxy (ULIRG) phase. HST images have shown that the host galaxy 
is in fact a post-merger remnant, and its spectrum shows evidence for multiple stages of star-formation as might be expected through such a process \citep{canalizo00a,
brotherton02,cales11}.

There are at least two possibilities to connect the presence of a post-starburst population and nuclear activity:
(1) the flow of gas towards the nucleus first triggers star formation in the circumnuclear region and the nuclear activity is triggered after hundreds of 
Myrs; in the meantime, the star formation may cease due to exhaustion of the gas \citep{sb01,davies07};
(2) the flow of gas towards the nucleus triggers star formation in the circumnuclear region and the nuclear activity, when triggered, quenches the star 
formation \citep{granato04,dimatteo05,hopkins06, canodiaz12}.
In the first scenario, the fueling of the black hole usually occurs with a delay of a few 100\,Myrs after the beginning of the star formation -- in order to 
explain the systematic excess of intermediate age stellar population contribution in AGN hosts, although there are a few cases in which simultaneous on-going 
star formation and nuclear activity are observed \citep{sb01}. In the scenario proposed by \cite{davies07} the gas fueling the nuclear activity originates 
from winds of AGB stars, which can be accreted more efficiently than the ejected mass by O and B stars and/or supernova. In these two versions of the first 
scenario, the star formation ceases due to the exhaustion of gas, and the post-starburst population is distributed in the circumnuclear region extending even 
to the nucleus.  
In the quenching scenario, the star formation occurring in the vicinity of the nucleus is abruptly interrupted due to feedback from the AGN (e.g., nuclear 
outflows, radiation and jets). Our proposition is that, in this case, the post-starburst population should be located in regions suffering the feedback 
effects -- e.g., over regions where a gaseous outflow is observed, which can be mapped via the gaseous kinematics.
With the goal of investigating the nature of the connection between the post-starburst stellar population and nuclear activity in PSQs, we began a program to map the 
stellar population and the manifestations of nuclear activity in the inner few kpc of PSQs using integral field spectroscopy.

In this paper we present the results obtained from observations of the PSQ SDSS J021011.5-090335.5 (hereafter PSQ J0210-0903), selected for its proximity, at a distance of 
only 170\,Mpc (from NED\footnote{The NASA/IPAC Extragalactic Database (NED) is operated by the Jet Propulsion Laboratory, California Institute of Technology, under contract 
with the National Aeronautics and Space Administration.}, for $H_\rmn{o} = 73.0 \pm 5$\,km\,sec$^{-1}$\,Mpc$^{-1}$), allowing the study of the spatial distribution of its 
stellar population and gas emission characteristics. At this distance, the scale at the PSQ is $\sim$ 0.83\,kpc\,arcsec$^{-1}$. The PSQ J0210-0903 is also one of the brightest 
PSQs with $z<$\,0.1, and its spectrum clearly reveals the presence of the Balmer jump and high-order absorption lines of the Balmer series. 
These absorption features are characteristic of the atmospheres of A stars, which dominate the absorption spectra of intermediate age ($\sim\,10^8$ yrs) 
stellar populations (or post-starburst galaxies).
The PSQ J0210-0903 is hosted by a barred spiral galaxy with Hubble type Sa \citep{graham09} and its SDSS Petrosian absolute magnitude is $M_\rmn{i} = -22.32 \pm 0.50$. 
Although this object is not as luminous as typical QSOs \citep[e.g., as those in][]{brotherton02,cales11} its absolute magnitude is close to the lower 
luminosity limit of QSOs ($M_\rmn{i} = -22.00$; \citealt{qsoDR7}) and its spectrum presents similar characteristics to other PSQs. In addition, the downsizing issue 
\citep{heckman04} makes the PSQs with lower redshifts important probes of AGN evolution.  

This paper is organized as follows. In Section \ref{sec:obs_reduc} we describe the observations and reduction processes. In Section \ref{sec:population} we present the 
methodology and results of the stellar population analysis. In Section \ref{sec:emitting} we present the methodology of analysis of the gas emission and the gas emission-line 
flux distributions and ratios. In Section \ref{sec:kinematics} we report our results for the gas kinematics. In Section \ref{sec:discussions} we discuss and interpret our 
results as well as discuss the technique of Principal Component Analysis to study the gas kinematics. In sections \ref{sec:mass} and \ref{sec:outflowrate} we present an estimate 
to the mass of the emitting gas and the mass outflow rate, respectively. In Section \ref{sec:conclusions} we present a summary of our results as well as our conclusions.

\begin{figure*}
\begin{minipage}[c]{1.0\textwidth}
\begin{center} 
  \includegraphics[scale=0.72,angle=270]{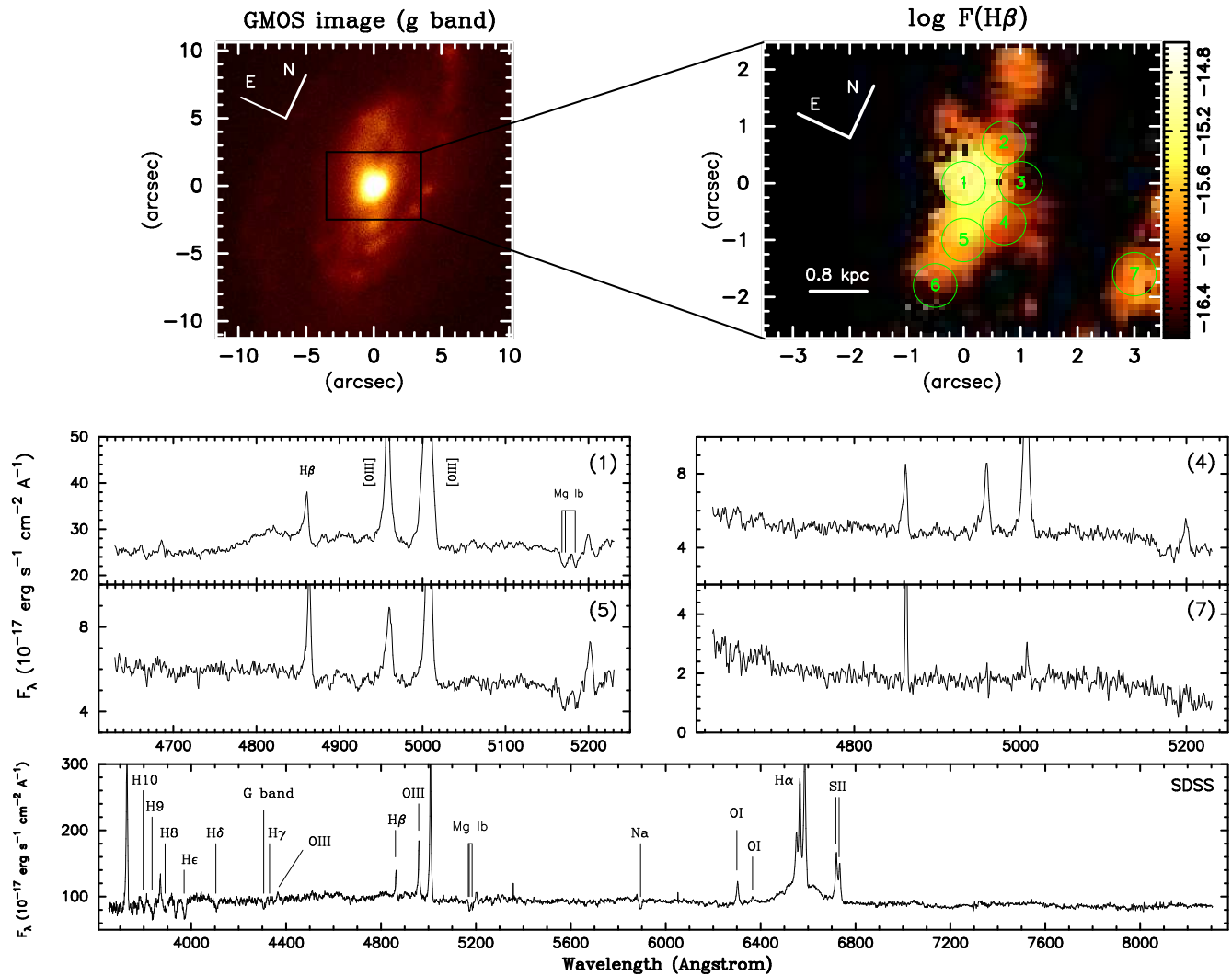}
  \caption{
           The top left-hand panel shows a GMOS g-band image of the PSQ J0210-0903 and a box that represents the field-of-view of the GMOS-IFU observations. The top 
	   right-hand panel shows the flux map in H$\beta$ with regions from which the spectra shown bellow were identified. Middle panels: spectra at regions 1, 4, 
           5 and 7 marked at the top right-hand panel showing the emission lines and that regions where H$\beta$ has a higher relativity intensity, 
	   the continuum is also bluer. Bottom: SDSS spectrum. The scale of the images is 830\,pc\,arcsec$^{-1}$.
           }
  \label{fig:galspec}
\end{center}
\end{minipage}
\end{figure*}

\begin{table*}
\begin{flushleft}
\begin{minipage}{174mm}
\centering
\caption{Synthesis results. Column 1: identification of spectra. Column 2: the spectral range, where $n$ means \textit{narrow} range (4630--5230\,\AA), $m$, 
         \textit{medium} range (4260--5260\,\AA) and $w$, \textit{wide} range (3780--8310\,\AA). Column 3: aperture radius of the spectra. Column 4: percent 
         contribution of the AGN featureless continuum to the flux at 4700\,\AA. Column 5-8: population vector in percentage of the flux at 4700\,\AA\
         for ages \textsc{young}, \textsc{intermediate} and \textsc{old}. The young population vector was separated in ionizing ($x_\rmn{Yi}$: $t\le5$\,Myrs) 
         and not ionizing ($x_\rmn{Ynoi}$: $t=25$\,Myrs) stars. Column 9-11: percent mass fraction for each age. Column 12: stellar mass enclosed in the 
         respective aperture in units of solar mass. Column 13: extinction in V-band magnitudes. Column 14: signal/noise ratio (SNR). Column 15: percent 
         difference between model and observed spectra.}
\label{tab:population}
\begin{tabular}{@{}ccccccccccccccc@{}}
\hline\hline
Position&$\Delta\lambda$& ap.      & $x_{\rm FC}$& $x_{\rm Yi}$& $x_{\rm Ynoi}$&$x_{\rm I}$&$x_{\rm O}$&$m_{\rm Y}$&$m_{\rm I}$&$m_{\rm O}$& $M$                  &$A_{\rm V}$& SNR  &adev  \\
        &               & (arcsec) & (\%)        & (\%)        & (\%)          &(\%)       &(\%)       &(\%)       &(\%)       &(\%)       & $\rmn{M}_\odot$      &(mag)      &      &(\%) \\
  (1)   &       (2)     &   (3)    & (4)         & (5)         & (6)           &(7)        &(8)        &(9)        & (10)      &(11)       &(12)                  & (13)      &(14)  & (15) \\
\hline\hline                       
SDSS    & $w$           &  1.5     & 32          & 18          & 0             & 13        & 36        & 0.18      & 0.86      & 98.95     & 1.2$\times 10^{10}$  & 0.282     & 38   & 2.0  \\
SDSS    & $m$           &  1.5     & 27          & 7           & 26            &  0        & 41        & 0.60      & 0.00      & 99.40     & 1.2$\times 10^{10}$  & 0.000     & 38   & 2.0  \\         
SDSS    & $n$           &  1.5     & 13          & 13          & 15            &  0        & 59        & 0.26      & 0.00      & 99.74     & 1.7$\times 10^{10}$  & 0.000     & 38   & 2.0  \\  
\hline
IFU 1   & $m$           & 1.5      & 15          & 5           & 29            &  0        & 51        & 0.91      & 0.01      & 99.08     & 9.1$\times 10^{9}$   & 0.000     & 39   & 1.9  \\
\hline 
IFU 1   & $m$           & 0.4      & 14          &  0          & 0             & 27        & 59        & 0.00      & 1.71      & 98.29     & 5.9$\times 10^{9}$   & 0.404     & 35   & 2.9  \\ 
IFU 2   & $n$           & 0.4      & --          & 45          & 0             &  1        & 54        & 0.23      & 0.26      & 99.50     & 1.3$\times 10^{9}$   & 0.152     & 27   & 2.8  \\                 
IFU 3   & $n$           & 0.4      & --          & 52          & 0             &  0        & 48        & 0.31      & 0.00      & 99.69     & 7.9$\times 10^{8}$   & 0.020     & 22   & 3.5  \\                 
IFU 4   & $n$           & 0.4      & --          & 58          & 0             & 12        & 29        & 0.56      & 1.37      & 98.06     & 4.5$\times 10^{8}$   & 0.000     & 20   & 4.0  \\                 
IFU 5   & $n$           & 0.4      & --          & 38          & 0             & 26        & 37        & 0.26      & 3.52      & 96.22     & 6.3$\times 10^{8}$   & 0.000     & 24   & 2.7  \\                 
IFU 6   & $n$           & 0.4      & --          &  0          & 1             & 99        &  0        & 0.07      & 99.93     & 0.00      & 6.4$\times 10^{7}$   & 0.000     & 13   & 6.1  \\                 
IFU 7   & $n$           & 0.4      & --          & 100         & 0             &  0        &  0        & 100.00    & 0.00      & 0.00      & 1.9$\times 10^{6}$   & 0.000     &  9   & 12.3 \\                 
\hline\hline                          
\end{tabular}                
\end{minipage}               
\end{flushleft}              
\end{table*}

\section{Observations and data reduction}
\label{sec:obs_reduc}

Two-dimensional spectroscopic data of the PSQ J0210-0903 were obtained in 2008 December 7 using the Gemini Multi-Object Spectrograph Integral Field Unit (GMOS-IFU) 
\citep{allington02}, in the program GN-2008B-Q-45. The observations were obtained in two-slit mode, covering a field of view (FOV) of 7.0\,$\times$\,5.2 arcsec$^2$, 
using the B600$\_$G5303 grating and g$\_$G0301 filter, resulting in a spectral range of $\sim$ $4400-5600$\,\AA\ and a wavelength sampling of 0.913\,\AA\ pixel$^{-1}$ at 
a velocity resolution R $\sim$ 3000 (FWHM\,$\sim$\,120 km\,s$^{-1}$). The total exposure time was 8$\times$1700 s. The seeing during the observation was approximately 
0\farcs6, corresponding to a spatial resolution at the distance of the galaxy of $\sim$\,0.5\,kpc.  

Data reduction was accomplished using generic {\small IRAF}\footnote{{\small IRAF} is distributed by the National Optical Astronomy Observatories, which is operated by the 
Association of Universities for Research in Astronomy, Inc. (AURA) under cooperative agreement with the National Science Foundation.} tasks and specific ones developed for 
GMOS data in the \emph{gemini.gmos} package. The basic reduction steps were trimming, bias subtraction, flat-fielding, cosmic rays cleaning, extraction of the spectra, 
sky-subtraction, wavelength and flux calibration, differential atmospheric refraction correction and coaddition of different exposures. As no standard stars were observed 
together with the galaxy the flux calibration was only relative. The absolute flux calibration was obtained from the ratio between a spectrum from the Sloan Digital Sky 
Survey (SDSS) database and our spectrum integrated within the same aperture (3\arcsec). This ratio was then used to scale the whole data cube. The final data cube contains 
3640 spectra each corresponding to an angular coverage of 0\farcs1\,$\times$\,0\farcs1 or 83\,$\times$\,83\,pc$^2$ at the distance of the galaxy. Cosmic 
rays were cleaned from the data before sky-subtraction with Laplacian cosmic ray identification routine {\small LACOSMIC} \citep{vandokkum01}. The spectra were corrected 
for reddening due to the interstellar Galactic medium using the {\small IRAF} routine \emph{noao.onedspec.deredden} for the $V$-band extinction $A_\rmn{V}=0.085$; its value 
was calculated using the NED extinction calculator, which uses the \cite{schlegel98} Galactic reddening maps.

In the top-left panel of Figure \ref{fig:galspec} we present the Gemini-GMOS $g$-band acquisition image of the PSQ, which we have rotated to the  same orientation of the 
IFU observations. In the top-right panel we present the H$\beta$ flux distribution within the IFU FOV, including the outline of a set of circular regions with 0\farcs8 
diameter from which we have integrated the spectra used in the stellar population synthesis (see next section). In the middle panels of the figure we present rest-frame 
spectra from the nucleus (the AGN, at region 1) together with those from two circumnuclear regions (regions 4 and 5) and from an H{\scriptsize II} region, at the more 
distant region 7. In the bottom panel the SDSS rest-frame spectrum shows the characteristic emission lines of the optical range, as well as the absorption lines from 
intermediate age stars, namely, the high-order Balmer lines between $3700-4000$\,\AA.            

\begin{figure*}
\centering
\begin{minipage}[c]{1.0\textwidth}
  \includegraphics[scale=0.35,angle=0]{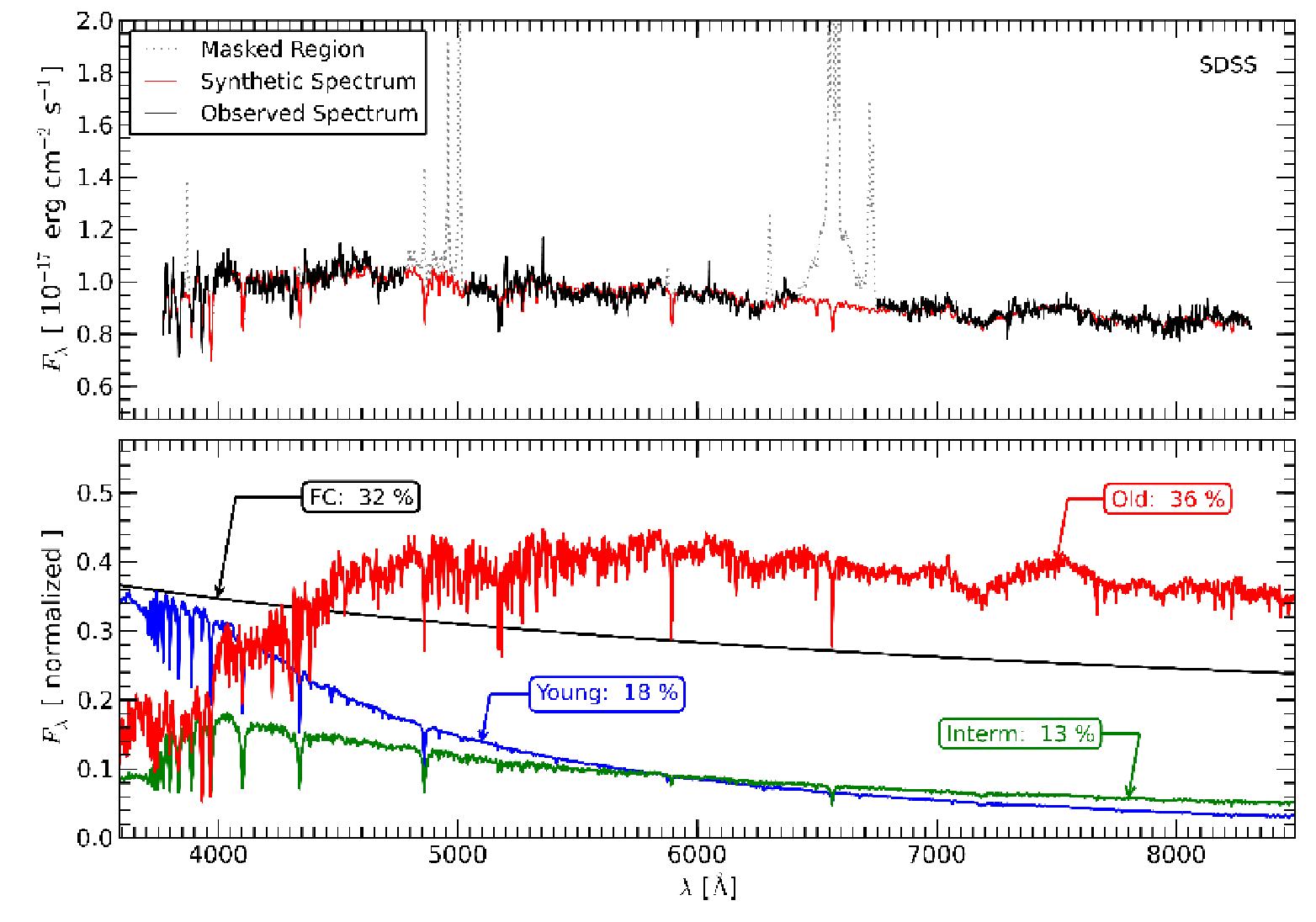}
  \includegraphics[scale=0.35,angle=0]{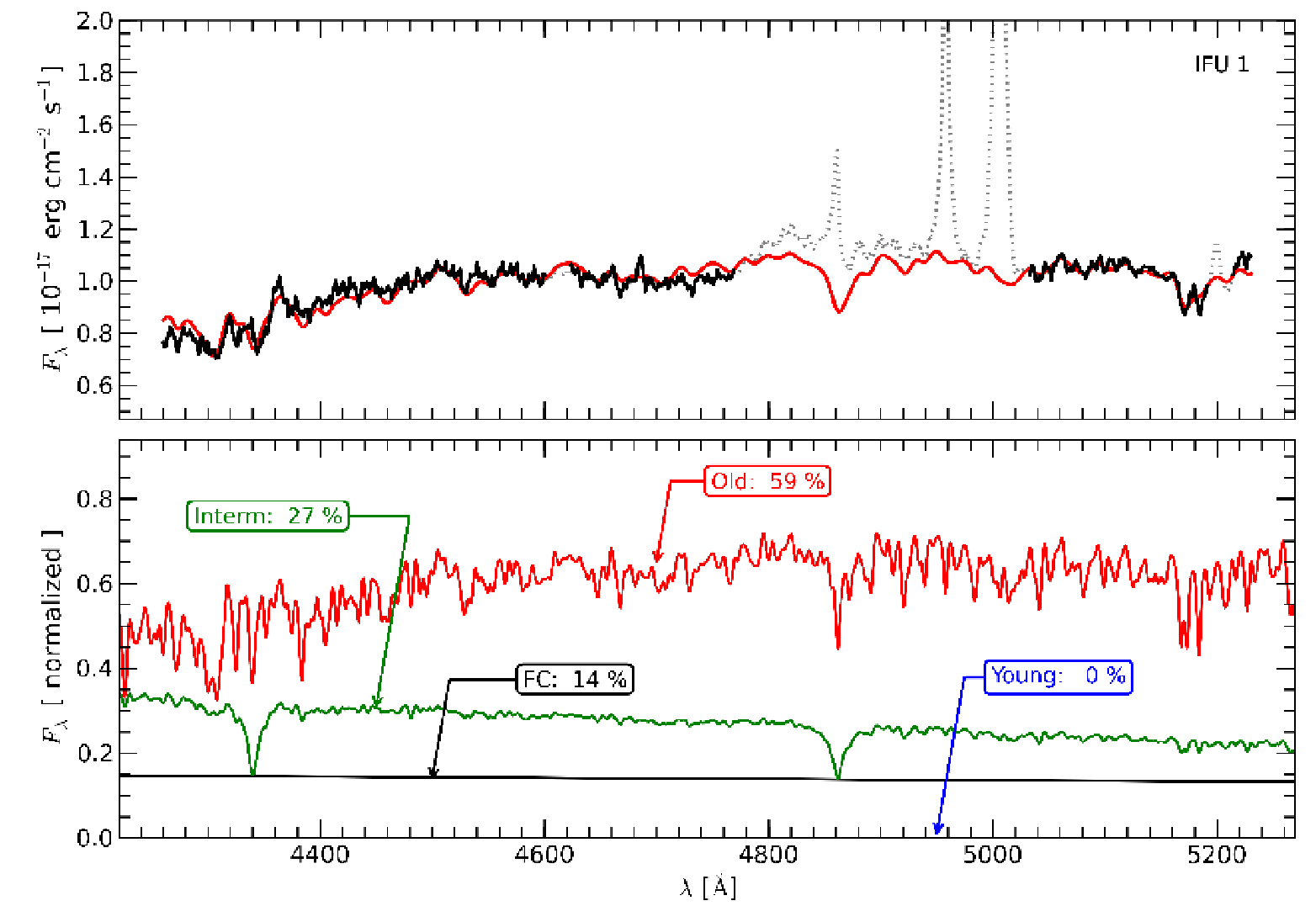}
  \includegraphics[scale=0.35,angle=0]{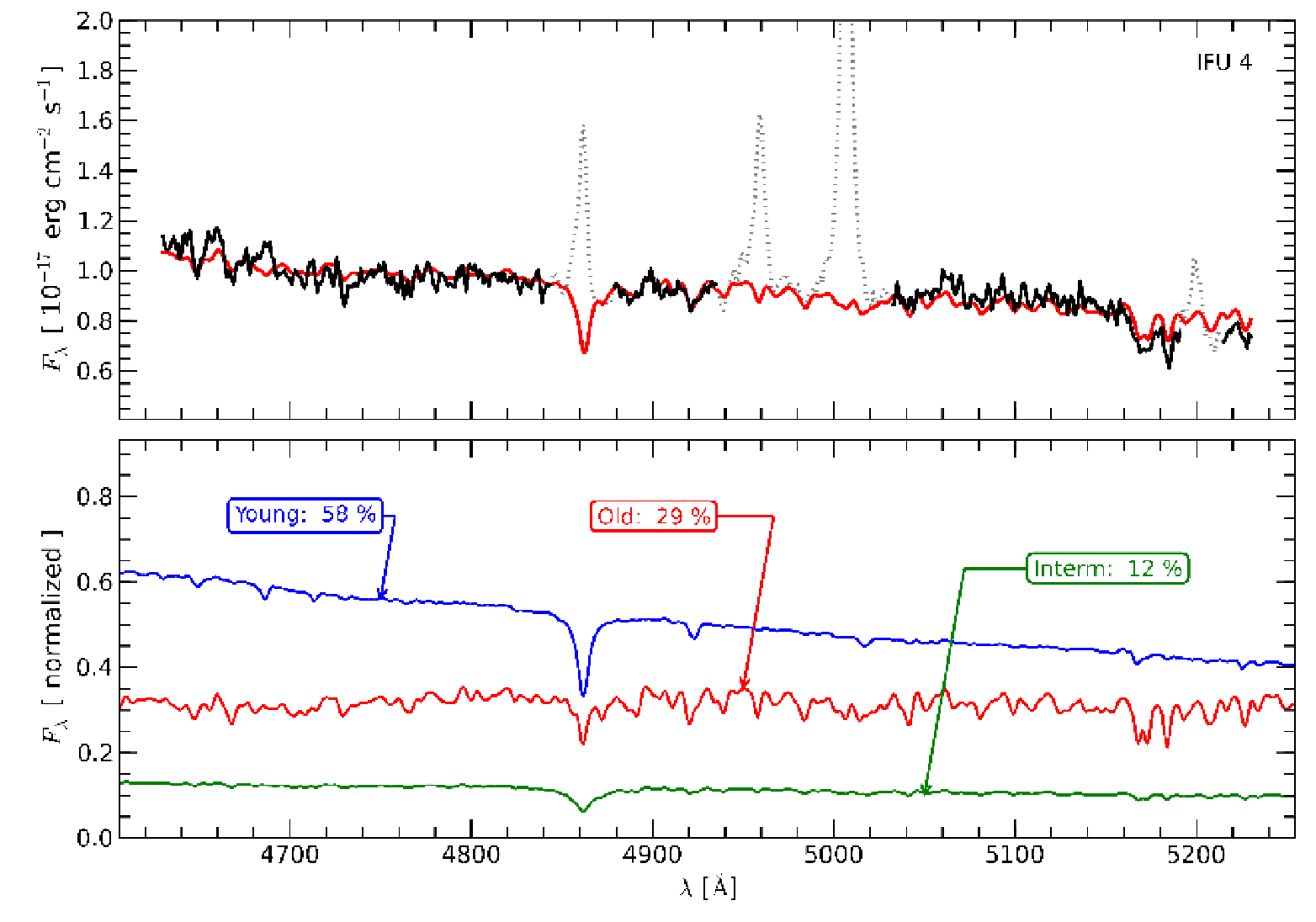}
  \includegraphics[scale=0.35,angle=0]{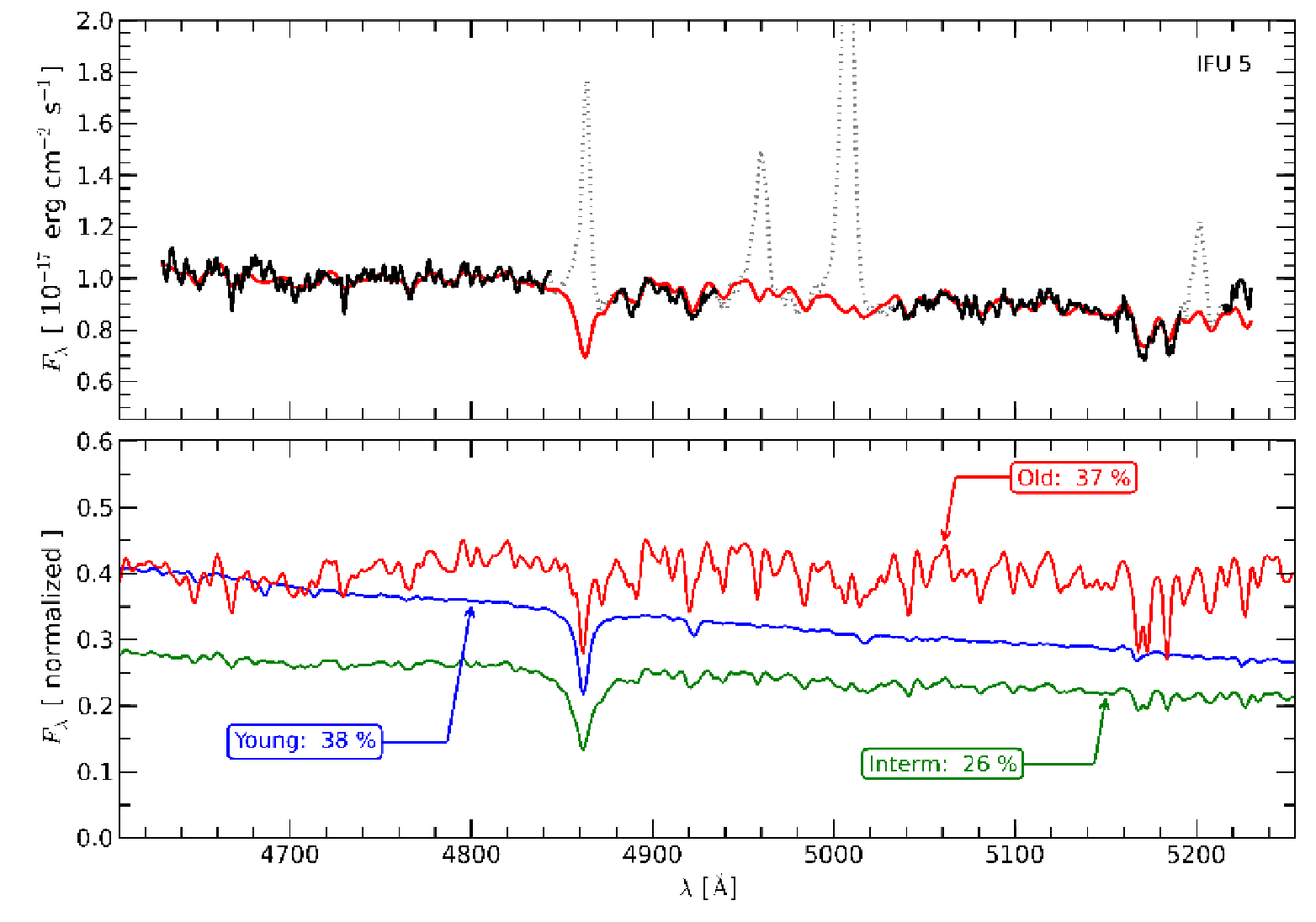}
  \includegraphics[scale=0.35,angle=0]{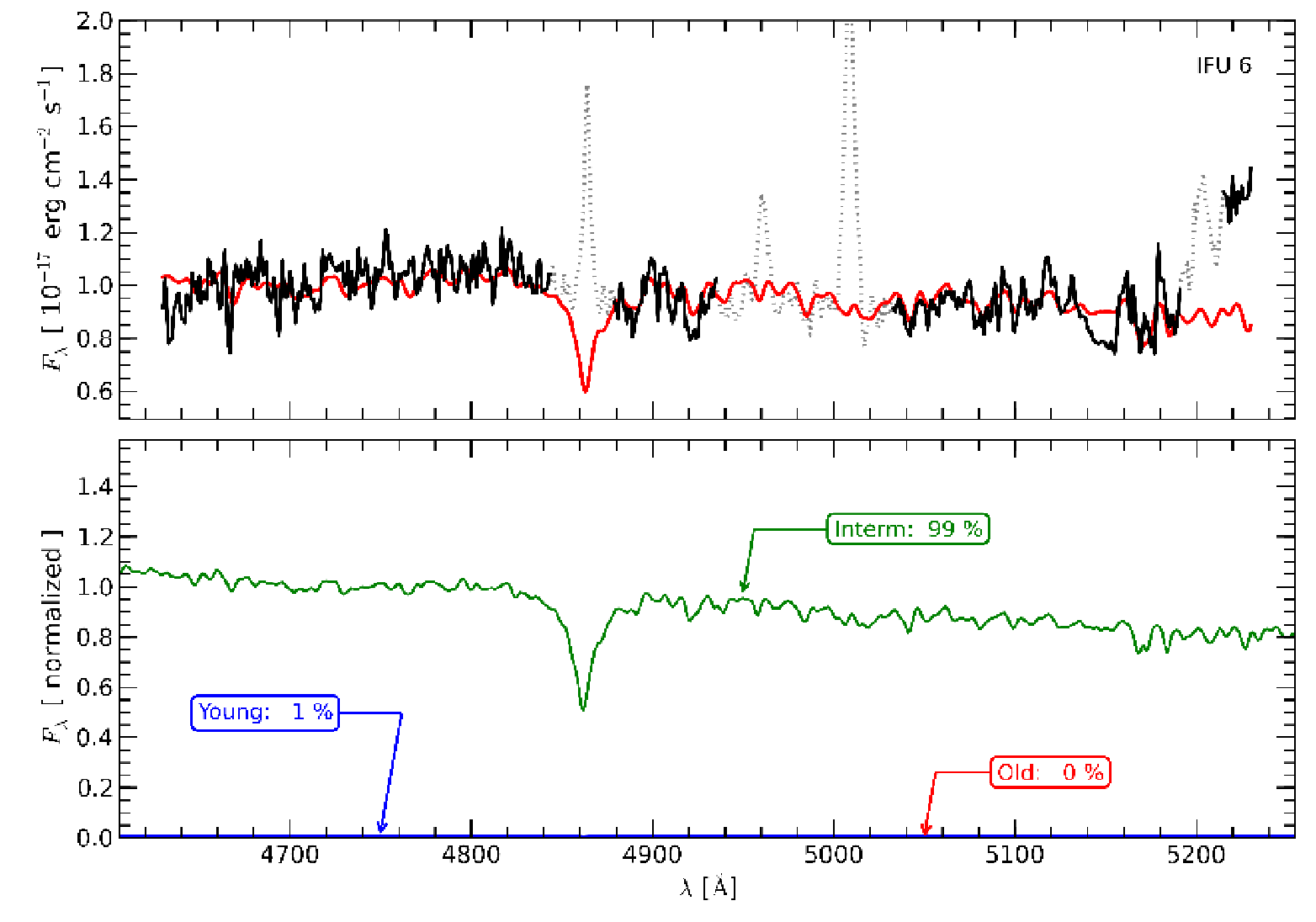}
  \includegraphics[scale=0.35,angle=0]{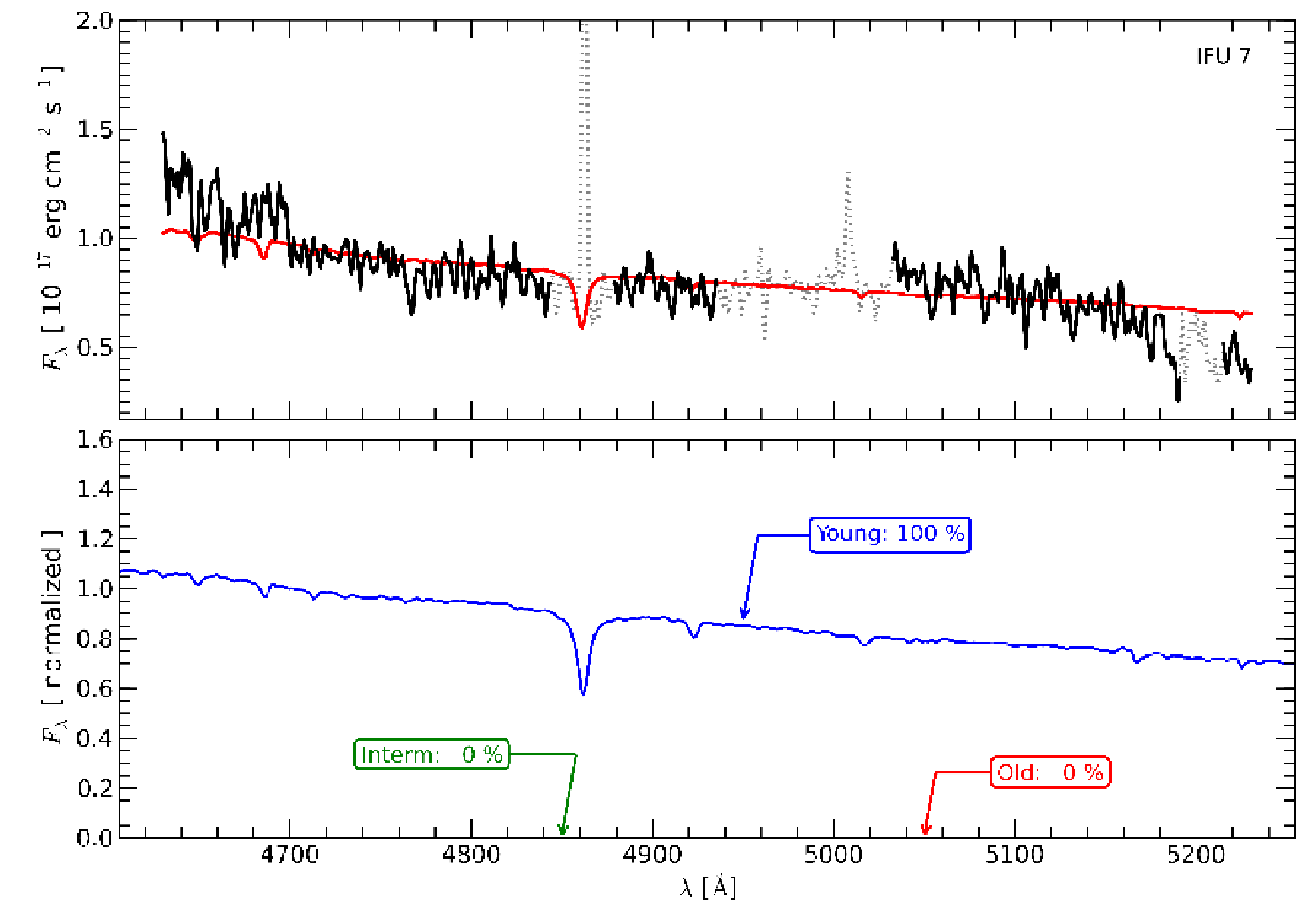}
    \caption{In this figure we present the synthesis results for the SDSS spectrum, for the nucleus (IFU 1), for the circumnuclear regions (IFU 4, 5 and 6) 
             and for the H\,{\scriptsize II} region located slightly farther from the nucleus (IFU 7). The black dashed-dotted line at the top of each panel 
             shows the observed spectrum. The red solid line shows the fitted model, while the dotted line shows the masked regions of the spectrum (emission lines or spurious 
             features). At the bottom we show the spectra corresponding to each age bin (\textsc{young} -- blue; \textsc{intermediate age} -- green; 
             \textsc{old} -- red) scaled to its percent contribution to the total continuum light at 4700\,\AA.}
\label{fig:obs+model}
\end{minipage}
\end{figure*}

\section{Stellar population}
\label{sec:population}
We have analyzed the stellar population using both the SDSS spectrum --- which corresponds to an aperture of 3\arcsec\ ($\sim$\,2.5\,kpc at the distance 
of the galaxy) -- covering the wavelength range $3770-8310$\,\AA\, and our integral field spectroscopy through the use of the technique of spectral synthesis.             

The Integral Field Spectroscopy could in principle be used to obtain information on the spatial distribution of the stellar population applying the synthesis to the 
spectra of each pixel. However, the signal-to-noise ratio (SNR) of individual spectra was not high enough, thus, in order to increase it, we integrated the spectra 
within the 0\farcs8 apertures described above and indicated in Figure \ref{fig:galspec}. These regions were selected as sampling extranuclear regions 
with high enough S/N ratio in the absorption spectra to allow performing spectral synthesis and being separated enough from the nucleus in order to avoid possible 
contamination from the nuclear light due to the PSF wings.

In order to perform the spectral synthesis we used the {\small STARLIGHT} code \citep{cid04,cid05,cid09}, which searches for the linear combination of $N_*$ simple 
stellar population (SSP), from a user-defined base, that best matches the observed spectrum. Basically, the code fits an observed spectrum $O_\lambda$ solving the 
following equation for a model spectrum $M_\lambda$ \citep{cid04}:

\begin{equation}
M_\lambda={M_{\lambda}}_{0} \left( {\sum_{j=1}^{N_*} {x_j}{b_{j,\lambda}}{r_\lambda}} \right) \otimes G(\upsilon_*,\sigma_*), 
\label{eq:M_lambda}
\end{equation}
where
${M_{\lambda}}_{0}$							is the synthetic flux at the normalization wavelength,
$\textbf{x}$								is the population vector, whose components represent the fractional contribution
									of the each SSP to the total synthetic flux at $\lambda_0$, 
$b_{j,\lambda} \equiv 
{L_{\lambda}^{SSP}(t_j,Z_j)}/{L_{\lambda_0}^{{SSP}}(t_j,Z_j)}$	is the spectrum of the $j$th SSP, with age $t_j$ and metallicity $Z_j$, normalized at $\lambda_{0}$,
$r_\lambda \equiv 10^{-0.4(A_\lambda-A_{\lambda_{0}})}$			is the reddening term, and 
$G(\upsilon_*,\sigma_*)$ 						is the Gaussian distribution, centred at velocity $\upsilon_*$ with dispersion $\sigma_*$, used to model 
									the line-of-sight stellar motions.
The reddening term is modeled by {\small STARLIGHT} due to foreground dust and parametrized by the V-band extinction $A_\rmn{V}$ so that all components are equally reddened
and to which we have adopted the Galactic extinction law of \cite{cardelli89} with $R_\rmn{V}=3.1$.

The fit of the model to the observed spectrum is carried out minimizing the following equation \citep{cid04}:
\begin{equation}
\chi^2= {\sum_{\lambda} \left[ (O_\lambda - M_\lambda) w_\lambda \right]^2}, 
\label{eq:qui2}
\end{equation}
where $w_\lambda$ is the weight spectrum, defined as the inverse of the noise in $O_\lambda$. Emission lines and spurious features are masked out by fixing $w_{\lambda}=0$ 
at the corresponding $\lambda$. The minimum of equation \ref{eq:qui2} corresponds to the best parameters of the model and the search for them is carried out with a simulated 
annealing plus Metropolis scheme. A detailed discussion of the Metropolis scheme applied to the stellar population synthesis can be found in \cite{cid01}.

We constructed the spectral base with the high spectral resolution evolutionary synthesis models of \cite{bc03} (BC03), where the SSPs cover 11 ages, $t=1.0\times10^6$, 
$5.0\times10^6$, $2.5\times10^7$, $1.0\times10^8$, $2.9\times10^8$, $6.4\times10^8$, $9.1\times10^8$, $1.4\times10^9$, $2.5\times10^9$, $5\times10^9$ and $1.1\times10^{10}$
\,yrs, assuming solar metallicity ($Z=0.02$). We have used the SSP spectra constructed from the STELIB library \citep{leborgne03}, Padova-1994 evolutionary tracks and 
\cite{chabrier03} Initial Mass Function (IMF). In order to account for the AGN featureless continuum (FC) a non-stellar component was also included, represented by a power 
law function ($F_\nu \propto \nu^{-1.5} $). In accordance to \cite{cid04} we have binned the contribution of the SSPs into three age ranges: \textsc{young} ($x_{\rm Y} 
\equiv x_{\rm Yi}+x_{\rm Ynoi}$, where $x_{\rm Yi}$ (t$\le$5\,Myrs) represents the ionizing population and $x_{\rm Ynoi}$ ($t=25$\,Myrs) represents the non-ionizing one), 
\textsc{intermediate} ($x_{\rm I}$ to $100\,\rmn{Myr} \leq t \leq 2.5\,\rmn{Gyr}$) and \textsc{old} ($x_{\rm O}$ to $t>2.5$\,Gyr). \cite{cid05} have shown 
that the star formation history of a galaxy may be very well recovered by this condensed population vector, as the individual contributions of each SSP are very uncertain. 
We have used this conservative spectral base, because we do not have sufficient constraints in our spectra to allow the use a broader spectral base. This procedure has been
extensively used in previous studies \citep[e.g., ][]{cid04,cid05} and the investigation of the impact of the corresponding assumptions in the fit of the observed spectra
is beyond the scope of the present paper. Our goal here is to investigate the relative contribution of broad age components as a function of location in the observed field.

The results of the synthesis are shown in Table \ref{tab:population} and Figure \ref{fig:obs+model}; they are presented in two sets, one for the SDSS spectrum and another for 
the IFU integrated spectra, which are discussed in the next two sections. 
\begin{figure*}
\begin{minipage}[c]{1.0\textwidth}
\centering 
\includegraphics[scale=0.75,angle=0]{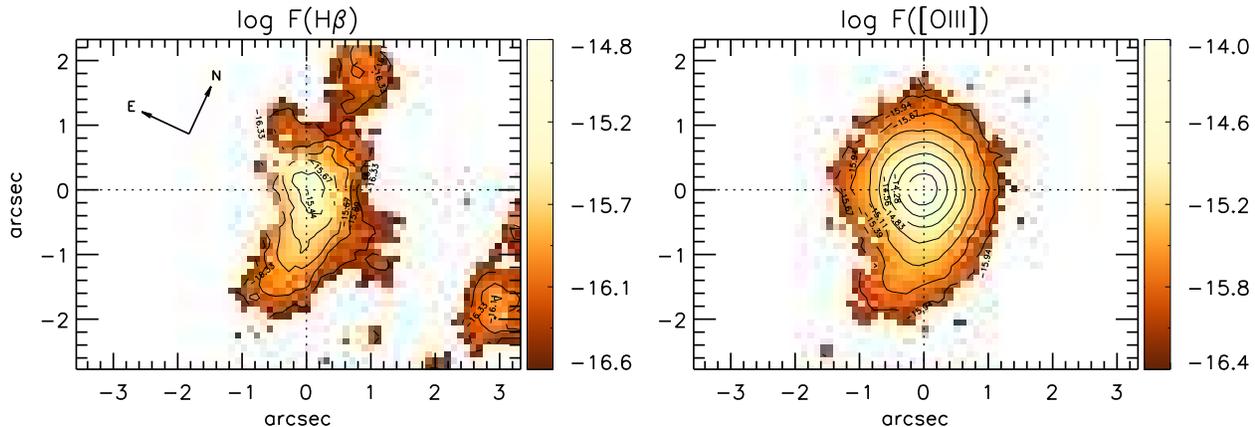}
\caption{
         Integrated fluxes in the \oiii\ and H$\beta$ emission line in logarithmic units of erg\,cm$^{-2}$\,s$^{-1}$, obtained from GH fitting using the 
         Levenberg-Marquardt least squares. The errors in the \oiii\ flux map are $\approx$\,4\% within the inner arcsecond, reaching up to $\approx$\,13\% 
         at the borders. In the H$\beta$ flux map the errors within the inner arcsecond are $\approx$\,11\%, increasing to $\approx$\,25\% at the borders of the mapped 
         field.
        }
 \label{fig:flux}
\end{minipage}
\end{figure*}

\subsection{Synthesis of the SDSS spectrum}
\label{sec:syntSloan}

The SDSS spectrum shows clear signatures of intermediate age stars which contribute approximately 13\% of the total luminosity at 4700\,\AA\ according to the synthesis 
(first line of Table \ref{tab:population}). The old population contributes with 36\% while the young and FC contribute with 18\% and 32\% respectively. 

Our IFU data cover a limited spectral range, from 4260 to 5230\,\AA, thus we investigate the effect on the synthesis of this restricted range using the SDSS spectrum. We 
have labeled as $m$ the spectrum covering this range in Table 1. For this spectral range the synthesis results show an increase in the contribution of the old 
population, which is 41\%, as well as of the young and FC components, which contribute with 33\% and 27\%, respectively, while the contribution of intermediate 
age stars disappears. Outside the nucleus, the blue end of the spectra are quite noisy, and we had to restrict the spectral range to the interval 
4630--5230\,\AA, which we have labeled $n$ in the Table \ref{tab:population}. In order to verify the effect of restricting the spectral range even further, we performed 
the synthesis of the SDSS spectrum also for this range. The result is a further increase of the old stellar population contribution to 59\%, and a decrease of the young 
population and the FC component, while the intermediate age population continues to be absent.

In summary, using the wider spectral range we find that the intermediate age population contributes with approximately 13\% of the flux at 4700\,\AA\ 
within the inner 3\,arcsec of our PSQ. However, when we restrict the spectral range to the intervals labeled as $m$ and $n$ we lose the sensitivity to the intermediate 
age stellar population, most probably due to the fact that the main signatures of intermediate age stars (high order Balmer absorption lines) have been removed. 

\subsection{Synthesis of the IFU spectra}
\label{sec:syntIFU}

We have first constructed an \qm{integrated} spectrum by adding the contribution of all spectra within the SDSS aperture (3\arcsec). We have then performed the spectral 
synthesis in order to check the consistency of our results. These results are shown in the fourth line of Table \ref{tab:population} for the \textit{m} wavelength interval, 
and they approximately agree with those for the SDSS, but presenting a difference of $\approx$10\% in the population vectors $x_\rmn{O}$ and $x_\rmn{FC}$. 
According to the \textit{adev} parameter, the percent difference between the model and the observed spectrum is $\approx$2\% for both the SDSS and the IFU spectra, which 
corresponds to a good fit. However, because the stellar population synthesis method does not provide error estimates for the population vector, we can use the difference 
of $\approx$10\% for the contribution of each component to the total light at 4700\,\AA\ as an error estimate for the population vector.
In the 5$^{\rm th}$ to the 12$^{\rm th}$ line of Table \ref{tab:population} we show the results of the synthesis for the set of circular regions described in section 
\ref{sec:obs_reduc} corresponding to apertures of 0\farcs8 diameter. As discussed above, our IFU data spectra cover a limited spectral range, namely the ranges labeled 
as $m$ and $n$. Only at the nucleus (region 1) we had enough SNR to cover the $m$ range, while in all other regions (2 to 7) we use the range $n$. 

In the nucleus (region 1), we have a predominance of the old population, contributing with 59\%, while the intermediate age population contributes with 27\% and the young 
+ FC contribute with 14\%. For the circumnuclear regions there is an increase in the contribution of the young populations, as we can see in the results 
for regions 2, 3, 4 and 5. The contribution of the old population decreases from 59\% --- at the nucleus --- to 54\%, 48\%, 29\% and 37\% respectively, showing that the 
old population has the largest contribution at the nucleus. At region 6, only the intermediate age component is present. The young and intermediate age stellar populations 
are thus distributed preferentially in the circumnuclear region, at about 0.8 kpc from the central AGN.
 
Figure \ref{fig:obs+model} shows the results of the synthesis for the SDSS spectrum and for five of the IFU integrated spectra of the circumnuclear 
region of the galaxy. The black dashed-dotted line, at the top of each panel, shows the observed spectrum. The red solid line shows the fitted model while the dotted 
line shows the masked regions of the spectrum. We have masked out emission lines and spurious features, such as the CCD gap. In each panel, at the bottom, we show the 
spectra corresponding to each age bin (identified as \textsc{young}, \textsc{intermediate age} and \textsc{old}) scaled to its percent contribution to the total 
continuum light at 4700\,\AA.

\section{Emitting Gas}
\label{sec:emitting}

\subsection{Measurement of the emission lines}
\label{sec:measureflux}

The kinematics and the flux distribution of the emitting gas in the central region of active galaxies have typically been measured using a single Gaussian 
function for each emission line profile. Nevertheless, a close inspection of the emission lines from the Narrow Line Region (NLR) of AGNs \cite[e.g.,][]{rogemar09a,rogemar10a}
shows that the profiles present deviations from a purely Gaussian profile, having \qm{blue} or \qm{red} wings, that are usually modeled by fitting a set of 
two or more Gaussians. In order to take into account the presence of these asymmetries, we used instead of Gaussians the Gauss-Hermite series (hereafter, GH),
with which we have fitted the emission-line profiles of H$\beta$ and \oiii\footnote{We denote the \oiii$\lambda5007$ emission line simply by \oiii}.
 
The GH series allows us to fit the wings of the emission lines via the $h_3$ and $h_4$ moments, in addition to the centroid velocities and the velocity dispersion. 
Following \cite{marel93} and \cite{cappellari04}, the GH series can be written as 
\begin{equation}
GH(w)=\gamma \frac{\alpha(w)}{\sigma} {\sum_{j=1}^{N} {h_j}{H_j(\omega)}}, \,\,\,\,\,w\equiv\frac{\lambda-\lambda_0}{\sigma},
\label{eq:GH}
\end{equation}
where $\gamma$ is the amplitude of the GH series, $\lambda_{\rm 0}$ is the central wavelength, $\sigma$ is the velocity dispersion, $h_j$ are the 
GH moments and $H_j$ are the Hermite polynomials and the function $\alpha(w)$ is the standard Gaussian, given by
\begin{equation}
\alpha(w)=\frac{1}{\sqrt{2\pi}}e^{-\frac{\omega^2}{2}}.
\end{equation}
For any choice of free parameters of the Gaussian ($\gamma$, $\lambda_0$ and $\sigma$) there is a GH series described by the equation (\ref{eq:GH}) that fits the emission 
line profile. A series that converge rapidly to the desired solution and that is very similar to the standard Gaussian function can be obtained by summing up to N=4 and taking 
the assumption that $h_0=H_0(w)=1$ and $h_1=h_2=0$ \citep{marel93}. In this case, the GH series given by equation (\ref{eq:GH}) can be approximated by 
\begin{equation}
GH(w)=\gamma \frac{\alpha(w)}{\sigma} \left[ 1+h_3H_3(w)+h_4H_4(w) \right], 
\label{eq:GH34}
\end{equation}
where $H_3(w)$ and $H_4(w)$ are the 3rd and 4th order Hermite polynomials.

The $h_3$ moment measures asymmetric deviations of a Gaussian profile, such as \qm{blue} ($h_3<0$) or \qm{red} ($h_3>0$) wings. The $h_4$ measures the 
symmetric deviations, where a wider and flatter profile than a Gaussian has $h_4<0$ and narrower and more peaked profile than a Gaussian has $h_4>0$. 
In general, the use of the GH series allows a better fit to any emission-line profile. 

The H$\beta$ and \oiii\ emission-line profiles were measured using an IDL routine that solves for the best solution of parameters of the GH series using the nonlinear
least-squares Levenberg--Marquardt method. The IDL routine used is similar to that detailed in \cite{profit10}, which makes use of the routine MPFIT\footnote{MPFIT 
provides a robust and relatively rapid way to run the least squares method and is available in http://purl.com/net/mpfit as part of \textit{Markwardt IDL Library}} 
\citep{markwardt09}, which performs the minimization. We have performed the fit to the spectrum of each pixel of the data cube in order to obtain the spatial 
distribution of the emission-line fluxes, velocity dispersions, centroid velocities and GH moments $h_3$ and $h_4$.

\subsection{Error estimates}
\label{sec:error}

Error estimates were obtained using Monte Carlo simulations adding to the original spectrum an artificial Gaussian noise normally-distributed, whose mean is equal 
to zero and the standard deviation equal to one. We have performed the Monte Carlo simulations with 200 realizations computing the best-fitting parameters of each 
realization so that we have the standard deviation of the best model in the end of the simulation. In principle, the larger the number of realizations, 
more reliable are the results, but in our case we have verified that with approximately 200 realizations the standard deviations are already of the same order as those
obtained with 500 to 1000 realizations. In this way, we have obtained the error estimates for the free parameters of the GH function: the centroid velocity, velocity dispersion, the 
moments $h_3$ and $h_4$ and the flux of the emission line. The error values in each measured parameter are given together with the description of the corresponding 
maps in the following sections.

\begin{figure}
\centering 
\includegraphics[scale=0.6,angle=0]{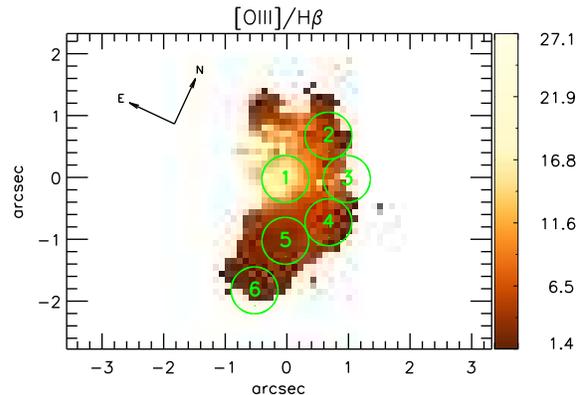}
\caption{
         Distribution of the emission-line ratio \oiii/H$\beta$. The regions from which spectra have been extracted for the study of the stellar population
         are identified by the circles.
        }
 \label{fig:ratio}
\end{figure}

\begin{figure*}
\begin{minipage}[c]{1.0\textwidth}
\centering 
\includegraphics[scale=0.75, angle=0]{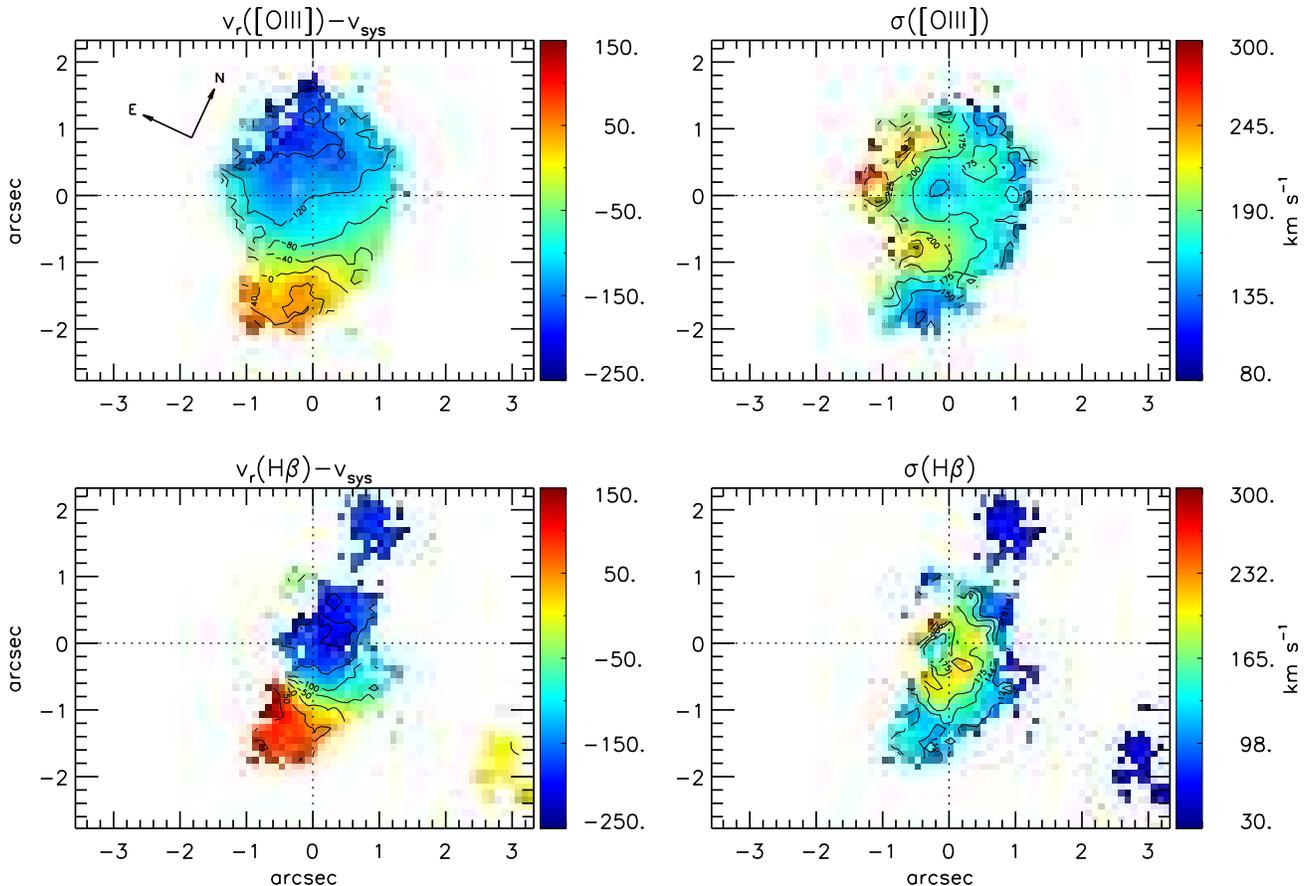}
 \caption{
          Left: Centroid velocity maps of \oiii\ and H$\beta$ in units of km\,s$^{-1}$. Right: Velocity dispersion maps. Both have been obtained from the fit of GH 
          series to the emission-line profiles. The mean uncertainties are between $\approx$6\,km\,s$^{-1}$ and 15\,km\,s$^{-1}$ for both the velocity 
         fields and velocity dispersion maps for \oiii. For H$\beta$ the mean uncertainties are between $\approx$10\,km\,s$^{-1}$ and 25\,km\,s$^{-1}$.
         }
 \label{fig:cinematica}
\end{minipage}
\end{figure*}

\subsection{Emission line flux distributions}
\label{sec:flux}
In Figure \ref{fig:flux} we present the flux distributions in the H$\beta$ and \oiii$\lambda$5007 emission lines, both given in logarithmic scale 
and erg cm$^{-2}$ s$^{-1}$ units. The location of the maximum brightness of the \oiii\ flux distribution coincides --- considering the spatial resolution 
of our data --- with the position of the peak of the continuum. This position was determined as the centroid of the flux distribution in the continuum, 
obtained by collapsing the data cube between two continuum wavelengths. We have adopted this position as the galaxy nucleus and its location in the figures
is identified either by two perpendicular dotted lines or a cross.

The H$\beta$ map shows a flux distribution elongated towards the South of the nucleus with a detached patch to the North. In the lower right corner there 
is another patch of emission which we attribute to an H{\scriptsize II} region located within what appears to be a spiral arm of the galaxy. The \oiii\ 
flux distribution is also elongated along the North-South, but along the East-West is broader and more symmetrically distributed around the 
nucleus than H$\beta$. \cite{graham09}, using SDSS images, have reported the presence of a bar along the North-South direction, where we have reported 
the elongations in \oiii\ and H$\beta$ flux distributions. The errors in the \oiii\ emission line fluxes are $\approx$\,4\% within the inner arcsecond, 
increasing to $\approx$\,13\% at the borders of the mapped field. In the case of H$\beta$ the errors within the inner one arcsecond are $\approx$\,11\%
while at the borders of the mapped field they reach $\approx$\,25\%.

In order to map the gas excitation, we show in Figure \ref{fig:ratio} the ratio between the \oiii\ and H$\beta$ flux maps, whose distribution presents the 
highest ratios --- and thus the highest excitation --- at the nucleus and up to $\approx$\,1\arcsec\ to the North, roughly within the circle (with diameter 
of 0\farcs8) labeled by the number 1. In this region the \oiii/H$\beta$ ratio reaches a value of $\ge$\,20, which decreases steeply outwards, down to values 
as low as $\approx$\,3, which is typical of LINERs and may include also contribution of ionization by young stars. The lowest values of \oiii/H$\beta$ are 
found at the regions labeled 4, 5 and 6, in regions where the stellar population synthesis results showed the largest contribution of young stellar populations.

\begin{figure*}
\begin{minipage}[c]{1.0\textwidth}
\centering 
\includegraphics[scale=0.75,angle=0]{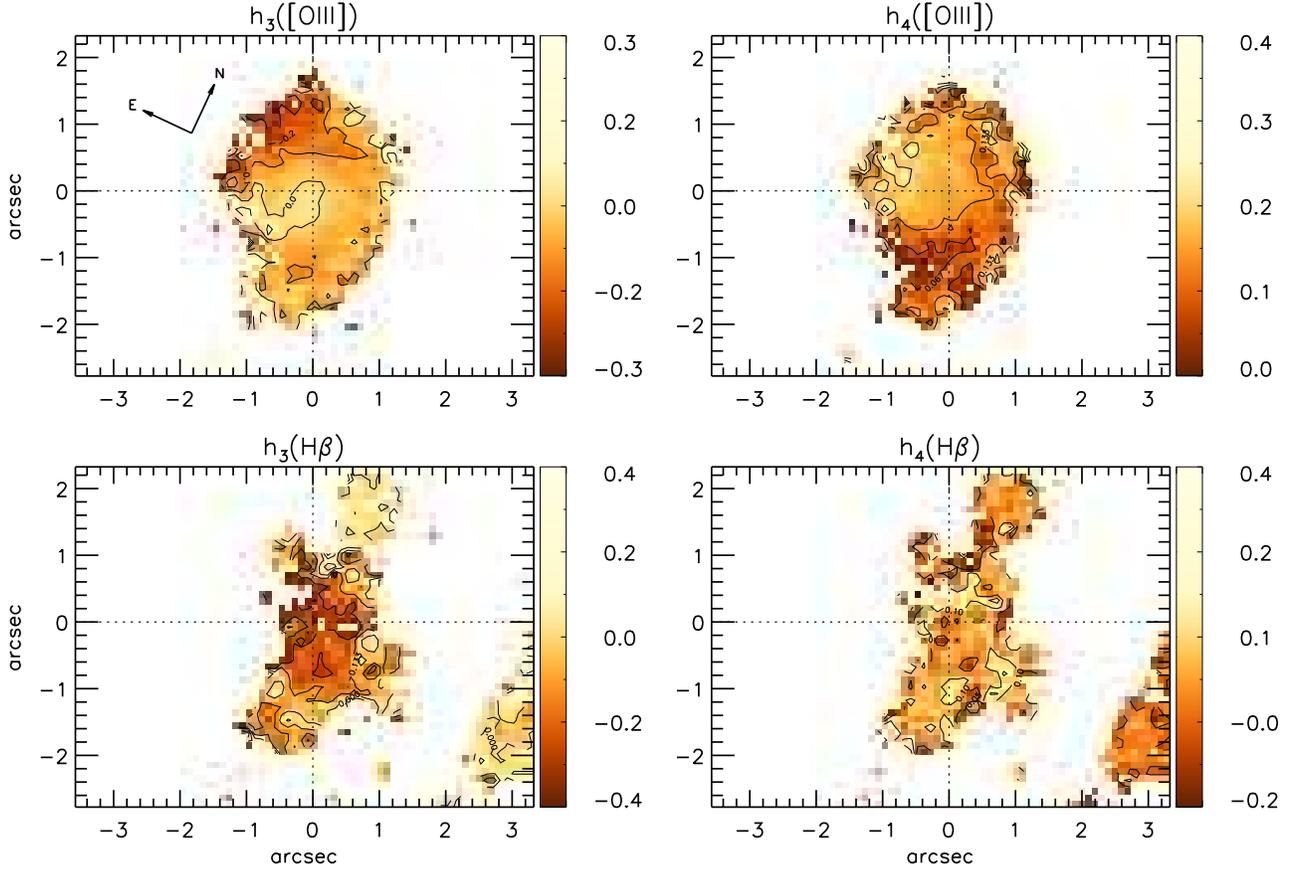}
 \caption{
          Distribution of the GH moments $h_3$ (left panels) and $h_4$ (right panels) of \oiii\ (top panels) and H$\beta$ (bottom panels) emission lines. 
          While the $h_3$ moment measures asymmetric deviations of a Gaussian profile, such as \qm{blue} ($h_3<0$) or \qm{red} ($h_3>0$) 
          wings, the $h_4$ measures the symmetric deviations, where a wider and flatter profile than a Gaussian has $h_4<0$ and a narrower and more peaked 
          profile than a Gaussian has $h_4>0$. The average errors for the $h_3$ and $h_4$ in \oiii\ are $\approx$\,0.02 within the inner arcsecond and 
          0.08 at the borders. The errors for $h_3$ and $h_4$ in the H$\beta$ are larger than for \oiii, increasing from 0.04 within the inner arcsecond
          to 0.2 at the borders.
         }
 \label{fig:momentos}
\end{minipage}
\end{figure*}

\section{Gas Kinematics}
\label{sec:kinematics}

As described in the previous section, we have obtained the gas kinematics --- centroid velocities, velocity dispersions and GH moments $h_3$ and $h_4$ --- using 
the GH series to fit the emission lines H$\beta$ and \oiii. In Figure \ref{fig:cinematica} we present, to the left, the centroid velocity maps and, to the right, 
the corresponding velocity dispersion maps. The velocity maps are masked according to the error estimates, which are on average in the range $10-20$\,km\,s$^{-1}$.
Pixels for which the error values are larger than these were masked. Centroid velocities are shown relative to the systemic velocity of the galaxy, 
12480$\pm20$\,km\,s$^{-1}$, whose value was measured from the velocity channel maps in the H$\beta$ emission-line, as explained in section \ref{sec:slices}.

\begin{figure*}
\begin{minipage}[t]{1.0\textwidth}
\centering 
  \includegraphics[scale=0.63,angle=0]{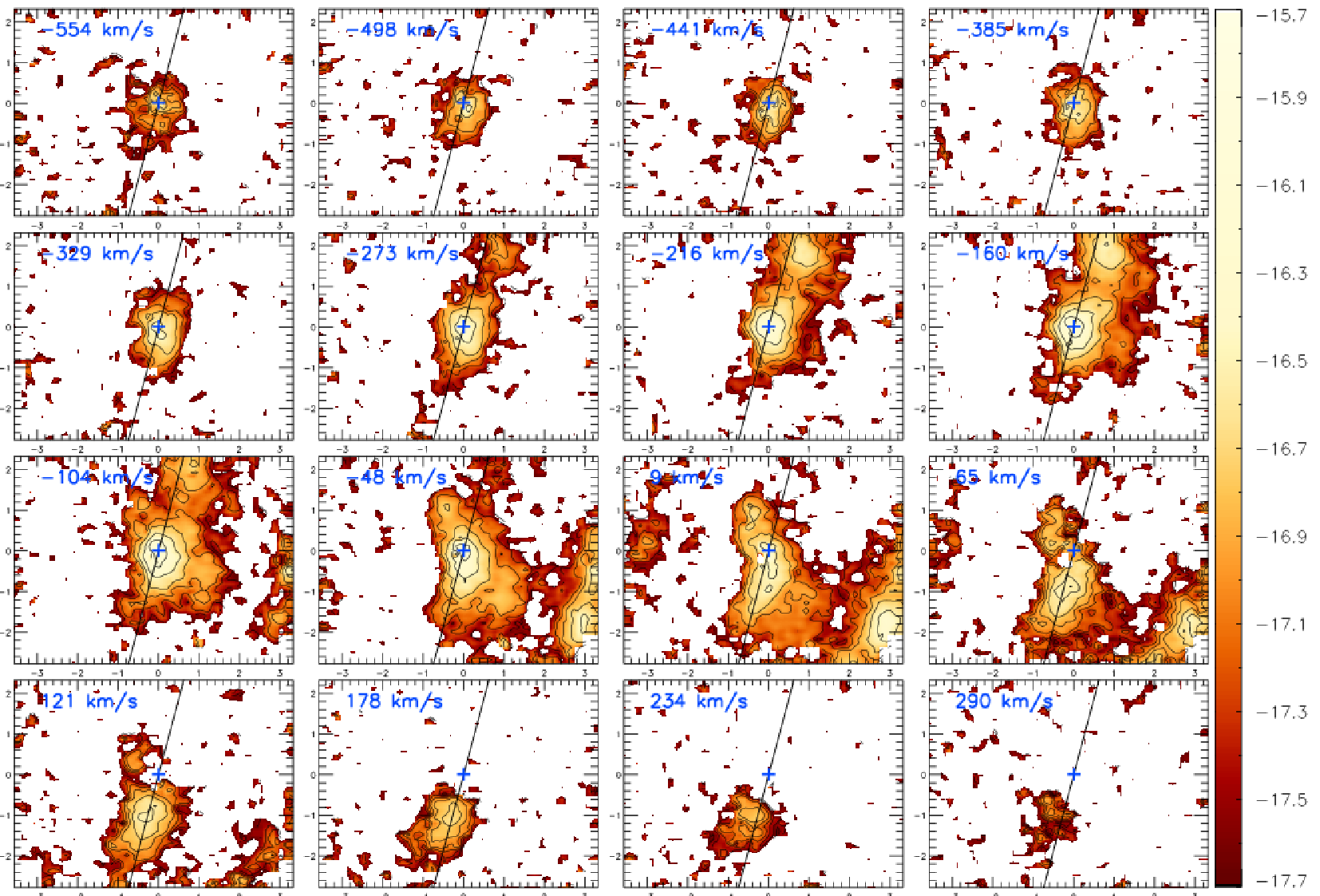}
   \caption{Channel maps obtained from slices -- of width with $\approx$\,55 km\,s$^{-1}$ -- along the emission-line profiles of H$\beta$. The numbers in the top left of 
            each panel are the central velocities of each slice in km\,s$^{-1}$. Fluxes are shown in logarithmic scale and in units of erg\,cm$^{-2}$\,s$^{-1}$. Spatial
            positions are in arcseconds. The solid black line represents the major axis of the galaxy.}
   \label{fig:sliceHb}
\end{minipage}
\end{figure*}

The \oiii\ velocity field (top left panel of Fig. \ref{fig:cinematica}) shows positive values towards the South and negative values towards the North, suggesting 
a rotation pattern with the line of nodes approximately along the North--South direction and a velocity amplitude of $\approx$\,150\,km\,s$^{-1}$. Alternatively, 
this pattern could also be due to a bi-polar outflow, frequently observed in the Narrow-Line Region of AGN. Considering that the velocity pattern is distorted 
relative to that corresponding to simple rotation and it's not centered at the nucleus, it is more likely that both rotation and outflows are present, as 
evidenced also by the blueshifts around the nucleus. The average errors in the velocity values increase from $\approx$\,6\,km\,s$^{-1}$ within the inner arcsecond 
to $\approx$\,15\,km\,s$^{-1}$ at the borders of the mapped field.

The \oiii\ velocity dispersion map (top right panel of Fig. \ref{fig:cinematica}) presents a range of values from $\approx$\,130\,km\,s$^{-1}$ to $\approx$\,300\,km\,s$^{-1}$. 
Within a radius of 0\farcs5 from the nucleus we observe the lowest $\sigma$'s, of $\approx$\,130\,km\,s$^{-1}$. This region is surrounded to the West by a semi-circle 
where the $\sigma$  values are slightly higher,  $\approx$\,175\,km\,s$^{-1}$, while to the SE, East and NE there is another semi-circle where the values reach 
even higher values of $\approx$\,250-300\,km\,s$^{-1}$. The errors and their distribution are similar to the ones for the \oiii\ velocity field.

The H$\beta$ velocity field (bottom left panel of Fig. \ref{fig:cinematica}) also presents a pattern that suggests rotation plus outflows with similar orientation 
to that observed in \oiii, but reaching a higher velocity amplitude of $\approx$\,200\,km\,s$^{-1}$. As in the case of \oiii, we also observe a rotational pattern
with kinematic center displaced from the continuum peak, showing blueshift at the nucleus. The average errors in the H$\beta$ velocity field range from 
10\,km\,s$^{-1}$ for the inner arcsecond to 25\,km\,s$^{-1}$ at the borders of the mapped field.

The H$\beta$ velocity dispersion map (bottom right panel of Fig. \ref{fig:cinematica}) shows a smaller region of 0\farcs2 near the the nucleus where the 
$\sigma$ has values of $\approx$\,130\,km\,s$^{-1}$ surrounded by a semi-circle to the West where the $\sigma$ is larger than in the case of \oiii\ (reaching 
$\approx$\,200\,km\,s$^{-1}$). Outwards, the values decrease to $\approx$\,100\,km\,s$^{-1}$. Although the H$\beta$ emission could not be measured as far as 
that of \oiii\ towards the East, the small region where the $\sigma(\rm{H}\beta)$ could be measured also shows an increase there, similar to what 
we have observed for $\sigma(\rm{\oiii})$.  

In the top panels of Fig. \ref{fig:momentos} we show the $h_3$ (left) and $h_4$ (right) GH moments of the \oiii\ emission-line profiles. The $h_3$ moment, 
over most of the IFU field, presents values between 0 and $-0.1$, indicating that there isn't significant asymmetric deviations with relation to Gaussian 
profiles. However, it reaches values of up to $-0.3$ to the NE, indicating the presence of blue wings in this region, giving additional support to the 
presence of outflows. The $h_4$ moment shows only positive values, meaning that the profiles are narrower and more \qm{pointy} than Gaussians. In the central 
region the $h_4$ moment has values around 0.2, increasing to 0.3 to the NE and decreasing to about 0.1 to the SW. The average errors obtained from the Monte 
Carlo simulations for the $h_3$ and $h_4$ are $\approx$\,0.02 within the inner arcsecond, increasing to 0.08 at the borders of the mapped field.
 
\begin{figure*}
\begin{minipage}[c]{1.0\textwidth}
\centering 
  \includegraphics[scale=0.63,angle=0]{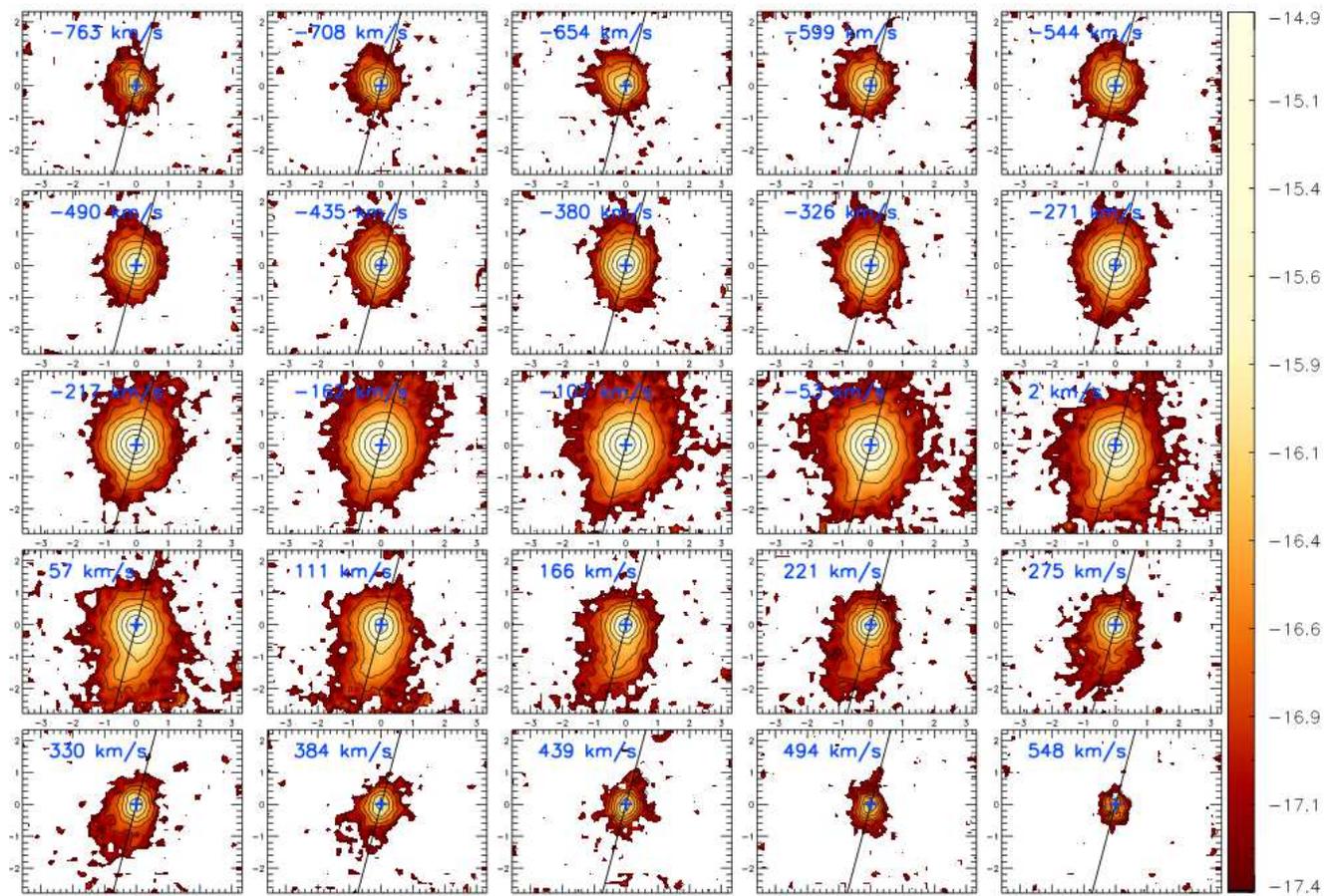}
   \caption{Channel maps obtained from slices -- of width $\approx$\,55 km\,s$^{-1}$ --  along the emission-line profiles of \oiii. The numbers in the top left of 
            each panel are the central velocities of each slice in km\,s$^{-1}$. Fluxes are shown in logarithmic scale and in units of erg\,cm$^{-2}$\,s$^{-1}$. Spatial
            positions are in arcseconds. The solid black line represents the major axis of the galaxy.}
   \label{fig:sliceOIII}
\end{minipage}
\end{figure*}

In the bottom panels of Fig. \ref{fig:momentos} we show the $h_3$ (left) and $h_4$ (right) GH moments of the H$\beta$ emission-line profiles. The $h_3$ moments 
present values of about $-0.3$ around  the nucleus, suggesting the presence of outflows towards us in this region, whereas in other regions the values are close to zero. 
The $h_4$ moment presents an irregular distribution, with values not deviating much from zero, indicating there are no significant symmetric deviations from a 
Gaussian profile for the H$\beta$ emission line. The average errors in $h_3$ and $h_4$ are larger than for \oiii, increasing from 0.04 within the inner arcsec 
to 0.2 towards the borders of the mapped field. 
 
\subsection{Channel maps}
\label{sec:slices}

We have obtained channel maps by integrating the flux distribution in different slices of velocity along the \oiii\ and H$\beta$ emission-line profiles. We have 
integrated the flux in each velocity channel after subtraction of the continuum contribution from both sides of the profile. This technique allows us to map the 
velocity field of the emitting gas throughout the emission line and not only at the central wavelength (like the velocity maps shown in Fig. \ref{fig:cinematica}).    

In Figure \ref{fig:sliceHb}, we present the channel maps of the H$\beta$ emission line for 16 velocities. The systemic velocity has been subtracted 
from the maps and its value was obtained as described below. The highest blueshifts, of $\approx\,-550$\,km\,s$^{-1}$, are observed at the nucleus. 
As the velocities become less negative (channel $-273$\,km\,s$^{-1}$), it is possible to observe emission outside the nucleus towards the North. 
For velocities around zero, we can observe emission distributed over most parts of the field, including in the spiral arm to the West of the nucleus and
in a similar structure to the East. As the velocities become more positive, the emission distribution moves to the South side. The highest positive velocities 
reach 290\,km\,s$^{-1}$ to the South of the nucleus. The kinematics observed in the velocity channels are consistent with a rotation pattern with line 
of nodes approximately along North-South and an outflow observed in negative velocities within $\sim$1 arcsec from the nucleus. As the nuclear outflow 
disturbs the rotational pattern in the nuclear region in the centroid velocity maps (Fig. \ref{fig:cinematica}), we have used the H$\beta$ velocity 
channels to estimate the systemic velocity, observing that the flux distribution pattern becomes symmetric relative to the nucleus along the galaxy
major axis for a systemic velocity of $12480\pm20$\,km\,s$^{-1}$. A comparison between the panels at $-273$ and $+290$\,km\,s$^{-1}$, for example, 
(approximately symmetric relative to the zero velocity) indeed shows that the extranuclear emission is approximately symmetric relative to the nucleus 
(within $\approx$\,20\,km\,s$^{-1}$). In addition, most of the extranuclear emission along the minor axis is found at zero velocity, supporting our adopted systemic velocity.

In Figure \ref{fig:sliceOIII}, we present the channel maps of the \oiii\ emission line in 25 different velocity channels, with an increase in velocity of 
$\approx$\,55\,km\,s$^{-1}$ from one panel to another. The highest negative velocities, that reach $-763$\,km\,s$^{-1}$ --- therefore higher than the ones observed 
for H$\beta$ --- are also observed at the nucleus, but extending circa 1\farcs0 towards Northeast (NE) of the nucleus. As the velocities become less negative, 
the emission becomes more extended toward the North. For velocities around zero we can observe the emission extending from North to South. As the velocities 
become more positive, the emission moves towards South, in the same way as observed for H$\beta$ (up to 275\,km\,s$^{-1}$). For \oiii\ even higher positive 
velocities are observed, however the peak values remain centered on the nucleus. \\

\section{Discussion}
\label{sec:discussions}

\subsection{Gas Excitation}
\label{sec:excitation}

We have investigated the gaseous excitation by mapping the ratio of \oiii/H$\beta$ and by measuring its value in the integrated spectra of the circumnuclear region, as
shown in column 6 of Table \ref{tab:avgage}. Additionally, we have separated the synthesis results for the young stellar population in terms of ionizing ($x_\rmn{Yi}$) 
and not ionizing population ($x_\rmn{Ynoi}$), given in percent contribution in columns 5-6 of Table \ref{tab:population}.

Using as reference the BPT diagnostic diagrams \citep{bpt81,kauffmann03,kewley06,stasinska06} we conclude that the emission-line ratios within the inner 
0.3\,kpc (radius) are typical of Seyfert nuclei with a mean value of \oiii/H$\beta=$17.9, but reaching up to $\sim$25 at locations closest to the nucleus. In the circumnuclear 
region smaller values are observed, decreasing from 8.1 (at region 2) to 0.2 in the detached region 7. At these locations (regions 2-7), the synthesis shows that the 
young population is essentially comprised by ages younger than 5\,Myrs. The BPT diagrams reveals additionally LINER-like emission-line ratios, which we have interpreted 
as due to excitation by the combined contribution of radiation from the AGN and hot stars.     

\begin{figure*}
\begin{minipage}[c]{1.0\textwidth}
\centering 
 \includegraphics[scale=0.73,angle=270]{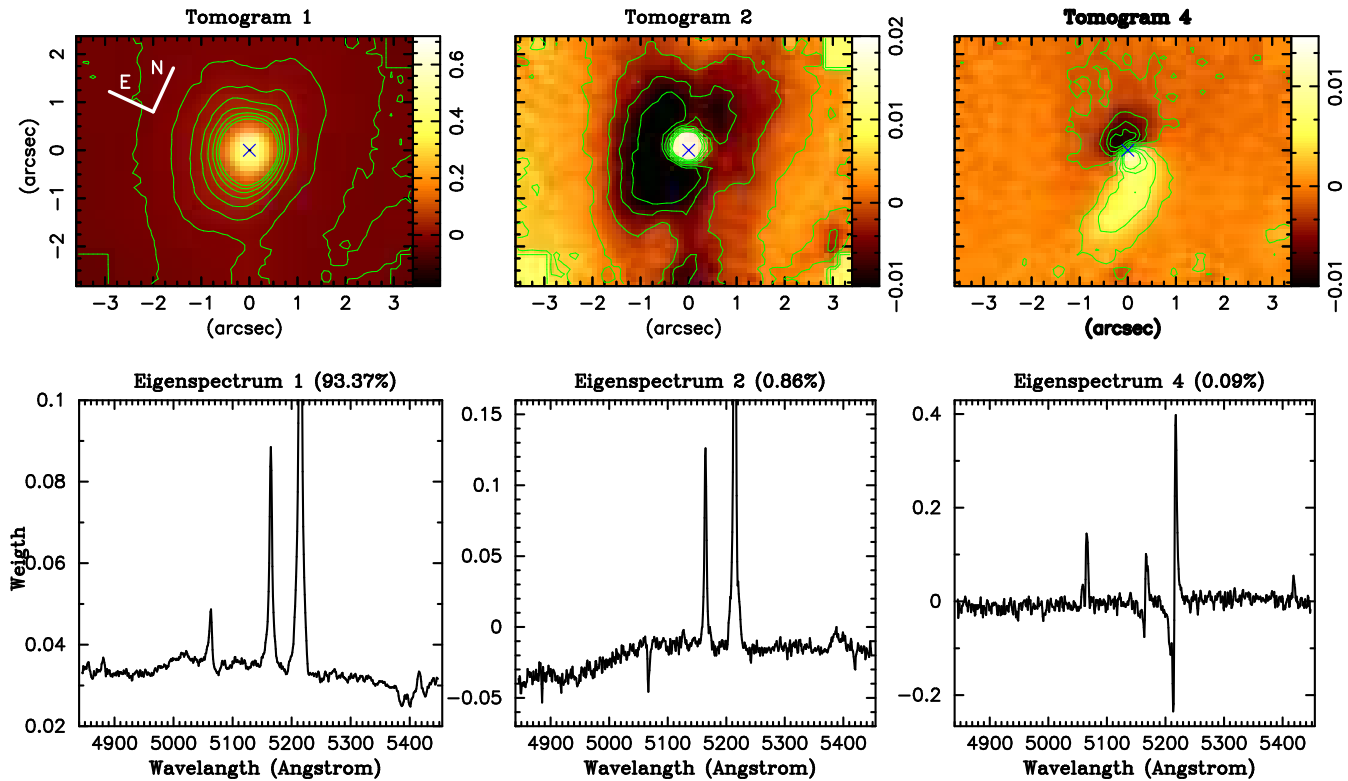}
  \caption{Tomograms 1, 2 and 4 of the PCA analysis (top) and their respective eigenspectra (bottom). The eigenspectra show the correlations and 
           anti-correlations in the data while the tomograms map them spatially. Positive and negative features in the eigenspectra are associated with regions in the
           tomograms whose color scales show respectively positive and negative features. The wavelength scale in the eigenspectra are given in the galaxy's reference 
           frame.}
 \label{fig:tomo}
\end{minipage}
\end{figure*}

\subsection{Gas Kinematics}
\label{sec:diss.kinematics}

The gaseous velocity fields of \oiii\ and H$\beta$ presented in Fig. \ref{fig:cinematica} both show similar rotation patterns, but with the rotation displaced 
from the continuum peak position by circa 1\farcs0 to the South. Such velocity field suggests that a possible rotation pattern around the nucleus is disturbed by 
a kinematic component not due to the gravitational potential of the bulge. In this region the emitting gas is blueshifted, with the velocity centroid reaching 
values of the order of $-200$\,km\,s$^{-1}$. The velocities observed in the wings of the emission-line profiles are, nevertheless, much higher, reaching $-760$\,km\,s$^{-1}$
for \oiii\ and $-670$\,km\,s$^{-1}$ for H$\beta$, as observed in the channel maps of Figs. \ref{fig:sliceOIII} and \ref{fig:sliceHb}. These blueshifts support the
presence of a nuclear outflow, in agreement also with the high velocity dispersions observed in this region (right panels of Fig. \ref{fig:cinematica}). Considering 
both the centroid velocities and the channel maps, the kinematic of the emitting gas can be understood as the combination of gas in rotation in the plane of the 
galaxy combined with an outflow component. 

The orientation of the rotation axis is approximately along the North-South direction, although the true orientation is not clear due the presence of the outflow 
in the central region. The rotation in the H$\beta$ emission is seen in the channel maps between $-273$ and $+290$\,km\,s$^{-1}$, with blueshifts observed along 
the major axis to the North of the nucleus. These blueshifts decrease gradually as the emission shifts towards the South side of the major axis and the velocity 
increases to redshifts. 

Additional support for the rotation in the \oiii\ centroid velocity field is given by the Principal Component Analysis of the data cube, which is described in the 
next section. The Principal Component 4 (PC4) together with its tomogram seems to be consistent with the presence of a rotational component, as shown in the channel
maps of the \oiii\ emission-line.

\subsection{PCA tomography}
\label{sec:pca}

Astronomical observations with IFU's provide a large amount of information, with two spatial dimensions and one spectral dimension, so that there are tens of millions 
of pixels to be analyzed. An efficient method to analyze all this information in an optimized way is the Principal Component Analysis (PCA), which has been adapted by 
\cite{steiner09} for application in IFU data.  The PCA technique has been previously employed in astronomy for morphological classification of galaxies \citep{lahav96} and for 
distinguishing radio-loud from radio-quiet QSOs \citep{boroson02}.

In order to extract additional information from our data and its complex gas kinematics, we have employed the PCA technique as described by \cite{steiner09}, 
to the calibrated data cube. The PCA allows us to separate the information initially contained in a system of correlated coordinates into a new system of uncorrelated 
coordinates, whose components are ordered in accordance with decreasing values of the variance of the principal components. The new coordinates are called 
\qm{eigenspectra} --- i.e., eigenvectors given as a function of wavelength --- so that their projection on the data produces images which are called 
\qm{tomograms}. While the eigenspectra show the correlations and anti-correlations in the data, the tomograms will map them spatially. The tomograms are 
like slices of data in the space of eigenspectra, whose interpretation can reveal correlations or anti-correlations between spectral and spatial 
characteristics of the data. 
  
In the panels of Figure \ref{fig:tomo} we present the results of the PCA applied to our data cube, where we show the principal components PC1, PC2 and PC4
(which show the largest variance of the data), represented through the corresponding tomograms and eigenspectra. We are not presenting the PC3, which has the 
third largest variance, because it contains mainly noise, mainly due to the difference in response between the blue and red slit detectors of the IFU. 
The other components do not present a significant variance. Correlations in the emission lines appear as all of them having positive values
in the eigenspectrum (e.g., bottom left panel of Fig. \ref{fig:tomo}). Anti-correlations appear as one line having positive value and the anti-correlated one having
negative value (e.g., H$\beta$ negative and \oiii\ positive in the bottom middle panel of Fig. \ref{fig:tomo}).

In the left panels of Figure \ref{fig:tomo} we show tomogram 1 and its eigenspectrum (PC1), which represents the principal component of the PCA 
with circa 93.4\% of the total variance. It shows a positive correlation between the emission lines, indicating that they are all generated 
in the same spatial region --- the brighter region of the tomogram. Since the emission lines that are present in the eigenspectrum have \oiii/H$\beta$ 
ratio typical of ionization by AGN, as well as contribution of the stellar population, we conclude that PC1 reveals the dominance of the AGN and the stellar population
over the entire central region.

The central panels of Figure \ref{fig:tomo} show eigenspectrum and tomogram 2, for which there exists an inverse correlation between \oiii\ and both the H$\beta$ 
emission and the stellar component, whose features are inverted and below zero in the eigenspectrum. The tomogram indeed shows that the corresponding 
emission originates in different regions: \oiii\ dominates in the inner $\approx$\,0\farcs3 (positive values), while the H$\beta$ and stellar population 
dominate in the surrounding region (negative values). The interpretation that we propose for PC2 is that although PC1 shows that most of the H$\beta$ 
emission is due to the AGN, there is some H$\beta$ emission which has a different origin. This emission is thus not generated by the AGN, but possibly 
by young stars that ionize the gas, indicating the presence of recent star formation around the nucleus. This conclusion is in agreement with the results 
of the stellar population synthesis, which show, in fact, the presence of young stars in these regions.

In the right panel of Figure \ref{fig:tomo} we show PC4, which represents 0.09\% of the variance. The eigenspectrum 4 shows that the \oiii\ emission 
has part of the profile in redshift and part in blueshift, and that they are anti-correlated, since the blueshift profile has negative values 
and the redshift profile has positive values; i.e. they originate in different regions, as seen in tomogram 4. The part in redshift originates 
in a compact structure to the Southwest of the nucleus which curves and extends to the South, while the blueshifted part originates in a compact 
region to the Northeast of the nucleus. This structure may be identified with the rotation component seen in the \oiii\ channel maps.

\subsection{Mass of the emitting gas}
\label{sec:mass}

We have estimated the total mass of ionized gas of the inner region (excluding the detached spiral arm) as:
\begin{equation}
M = V \epsilon\,n_{\rm e} m_{\rm p},
\label{eq:mass0}
\end{equation}
where $n_{\rm e}$ is the electron density, $m_{\rm p}$ is the proton mass, $\epsilon$ is the filling factor and $V$ is the volume of the emitting region.
 
The filling factor can be calculated using the H$\beta$ luminosity, as follows. The emissivity of the H$\beta$ line is given by \citep{osterbrock06}:
\begin{equation}
j_{\rm H\beta}=n_{\rm e} n_{\rm p}{\alpha^{\rm eff}_{H\beta}}\frac{h\nu_{\rm H\beta}}{4\pi}\,, 
\end{equation}
where $n_{\rm p}$ is the proton density, $\alpha_{\rm H\beta}^{eff}$ is the H$\beta$ effective recombination coefficient and $h\nu_{{\rm H\beta}}$ is 
the corresponding energy. The total luminosity in H$\beta$ is given by integrating the emissivity over the entire emitting volume and over all directions,
assuming a pure hydrogen gas which is completely ionized ($n_{\rm e} \equiv n_{\rm p}$):
\begin{equation}
L(\rmn{H}\beta) = \int \mskip-8mu \int j_{\rmn{H}\beta}\,\,\rmn{d}\bmath{\Omega}\rmn{d}V = 1.24 \times 10^{-25}\,V\,\epsilon\,n_e^2 \,\,\,\rmn{erg\,s}^{-1}\,,
\label{eq:intL}
\end{equation}
where we have adopted the $\alpha_{\rm H\beta}^{eff}$ for a temperature of $10^{4}$\,K. From the above equation, we obtain the product $V\epsilon$:
\begin{equation}
V\epsilon = 8.1 \times 10^{59}\dfrac{L_{41}({\rm H}\beta)}{n_3^2}\,\,\,\rmn{cm}^{-3},
\label{eq:epsilon}
\end{equation}
where $L_{\rm 41}({\rm H\beta})$ is the H$\beta$ luminosity, in units of $10^{41}$\,ergs\,s$^{-1}$ and $n_3$ is the electron density in units 
of $10^3$\,cm$^{-3}$. Thus, we can estimate the mass of the emitting region by inserting the expression \ref{eq:epsilon} for $V\epsilon$ above
in \ref{eq:mass0}, obtaining:
\begin{equation}
M\approx7\times10^{5}\dfrac{L_{\rm 41}({\rm H\beta})}{n_3}\,\,\,M_\odot\,,
\label{eq:mass1}
\end{equation}
given in units of solar masses \citep{peterson97}. 

The H$\beta$ luminosity was calculated from the integrated H$\beta$ flux F(H$\beta$) of the inner region corrected for the reddening obtained 
from the SDSS spectrum ($C({\rm H\beta}) = 1.38 \pm 0.05$), using the reddening law of \cite{cardelli89} adopting the theoretical 
$\frac{F(\rmn{H}\alpha)}{F(\rmn{H}\beta)}$ ratio of 3.0, corresponding to case B recombination \citep{osterbrock06}. For the assumed  distance 
of $d=$\,170\,Mpc, we obtain $L({\rm H\beta})=4\pi d^2 F({\rm H}\beta)10^{C({\rm H}\beta)} = 1.79\pm0.21 \times 10^{41}$\,ergs\,s$^{-1}$. 
Using the [S\,{\scriptsize II}] \,$\lambda6716/\lambda6731$ ratio, obtained from the SDSS spectrum, the electron density $n_3$ was obtained by solving 
numerically the equilibrium equation for a $5$-level atom using the {\small IRAF} routine \textsc{stsdas.analysis.nebular.temden} \citep{derobertis85,shaw94}.
The assumed electron temperature was $16000$\,K \citep{peterson97}, resulting in $n_3=0.150\pm0.02$. With these assumptions, we obtain a mass of ionized gas of 
$8.34\pm0.92 \times 10^6 M_\odot$. 

\subsection{Mass outflow rate}
\label{sec:outflowrate}

We have concluded from the gas kinematics that the blueshifted emission in the nuclear region is due to outflowing gas. Although the geometry of the 
outflow is not clear from our maps, previous kinematic studies \citep[e.g.,][]{crenshaw00,das05,das06,sb10} of other active galaxies show that AGN 
outflows present a conical geometry. On the basis of these previous studies, we assume that the outflow in PSQ J0210-0903 also has a conical geometry,
with the cone axis directed approximately towards us. We thus consider that the outflowing gas is crossing the base of a cone whose radius we estimate
from the extent of the blueshifted region around the nucleus seen in the channel maps. We have measured the ionized gas mass outflow rate at negative 
velocities through a circular cross section with radius $r=0$\farcs6 around the nucleus, assuming that the height of the cone is equal to the diameter 
of its base. The mass outflow rate can be calculated as \citep{rogemar11a,rogemar11b}:
\begin{equation}
\dot{M}_\rmn{out}=m_\rmn{p} n_{e} v_\rmn{out} A \epsilon \,,
\end{equation}
where $A=\pi r^2 \approx 7.42 \times 10^{42}$\,cm$^2$ is the area of the circular cross section, $m_\rmn{p}=1.67 \times 10^{-24}$\,g is the proton mass,
$n_\rmn{e}$ is the electron density and $\epsilon$ is the filling factor. The filling factor can be obtained from equation \ref{eq:epsilon} above taking 
into account only the luminosity of the gas in outflow. We have considered that the outflowing gas corresponds to that observed in the velocity channels 
between $-667$ and $-104$\,km\,s$^{-1}$. The total flux was obtained by adding the flux of the channels centered between $-667$ and $-329$\,km\,s$^{-1}$ 
and by adding 50\% of the flux of the channels between $-273$ and $-104$\,km\,s$^{-1}$, for which the remaining emission -- the other 50\% -- is assumed, 
as a best guess, to originate in the plane of the galaxy. The total flux in these channels is $F(\rmn{H}\beta) = 1.68 \times 10^{-14}$\,erg\,cm$^{-2}$\,s$^{-1}$ 
and $L(\rmn{H}\beta)=4\pi d^2 F(\rmn{H}\beta)=5.81\times 10^{40}$ \,erg\,s$^{-1}$. From the adopted geometry, the volume of the cone whose height $h$ 
and radius $r$ are such that $h = 2\,r = 996$\,pc is $V=7.6 \times 10^{63}$\,cm$^3$. The velocity adopted for the outflow is the average of the velocities 
of each channel above weighted by the corresponding flux, which results $v_\rmn{out} = -340$\,km\,s$^{-1}$. Adopting $n_e=150$\,cm$^{-3}$ (obtained from the 
[S{\scriptsize II}] line ratio from the SDSS spectrum), $\epsilon = 0.0027$, we obtain $\dot{M}_\rmn{out} = 1.1$\,M$_\odot$\,yr$^{-1}$. For a somewhat larger 
gas density, of $n_e=500$\,cm$^{-3}$ (as the aperture of the SDSS is too large, and the inner gas density being probably larger), we obtain $\epsilon= 0.00025$ 
and $\dot{M}_\rmn{out}=0.3$\,M$_\odot$\,yr$^{-1}$. 

\cite{rogemar11b} have obtained $\dot{M}_\rmn{out}=6$\,M$_\odot$\,yr$^{-1}$ for Mrk\,1157. They have also revised previous results for another 6 active 
galaxies in \cite{barbosa09} and in \cite{rogemar09b}, obtaining values for the mass outflow rate in range of $\dot{M}_\rmn{out} \approx 0.1-6$\,M$_\odot$\,yr$^{-1}$. 
Similar values, of $\approx 2.0$\,M$_\odot$\,yr$^{-1}$, have been reported by \cite{sb10} for NGC\,4151 as well as by \cite{veilleux05}, with $\dot{M}_\rmn{out} \approx 
0.1-10$\,M$_\odot$\,yr$^{-1}$ for a set of luminous active galaxies. The mass outflow rate that we have estimated for PSQ J0210-0903 is thus in good agreement with 
previous estimates for other AGNs.

Now we can compare the outflow mass rate with the accretion rate necessary to feed the AGN, which can be calculated as \citep{peterson97}:
\begin{equation}
\dot{m}=\dfrac{L_\rmn{bol}}{c^2\,\eta} \approx 1.8 \times 10^{-3} \left( \dfrac{L_{44}}{\eta}\right) \,\,{\rm M_\odot}\,\rm{yr^{-1}} \,,
\label{eq:mdot}
\end{equation}
where $L_\rmn{bol}$ is the nuclear bolometric luminosity, $c$ is the light speed and $\eta$ is the efficiency of conversion of the rest mass energy of the accreted material 
into radiation power and $L_{44}$ is bolometric luminosity in units of 10$^{44}$\,erg\,s$^{-1}$. Following \cite{rogemar11a,rogemar11b}, $L_\rmn{bol}$ can be 
approximated as $\approx\,100\,L(\rmn{H}\alpha)$, where $L(\rmn{H}\alpha)$ is the H$\alpha$ nuclear luminosity. Using the interstellar extinction coefficient $C(\rmn{H}\beta)$, 
the intrinsic ratio $\frac{F(\rmn{H}\alpha)}{F(\rmn{H}\alpha)}=3.0$ \citep{osterbrock06} and the reddening law of \cite{cardelli89}, we have obtained $F(\rmn{H}\alpha) = 1.40 
\times 10^{-13}$\,erg\,cm$^{-2}$\,s$^{-1}$ within 0\farcs4 of the nucleus. At the distance of 170\,Mpc, $L(\rmn{H}\alpha) = 4.84 \times 10^{41}$\,erg\,s$^{-1}$ and the 
corresponding nuclear bolometric luminosity is $L_\rmn{bol} = 4.84 \times 10^{43}$\,erg\,s$^{-1}$. If we assume an efficiency $\eta \approx 0.1$, for an optically thick 
and geometrically thin accretion disc \citep{frank02} we derive an accretion rate of $\dot{m} = 8.7 \times 10^{-3}\,\rmn{M_\odot}$\,yr$^{-1}$. Since $\dot{M}_\rmn{out}$ 
is 2 orders of magnitude higher than $\dot{m}$, we conclude that the outflowing gas does not originate only from the AGN, but is composed mostly of interstellar gas from 
the surrounding region of the galaxy, which is swept away by the AGN outflow. 

We can additionally estimate the kinetic power of the outflow, considering both the radial and turbulent component of the the gas motion. Using the H$\beta$ kinematics
to estimate this power, we have
\begin{equation}
\dot{E}_\rmn{out} \approx \frac{\dot{M}_\rmn{out}}{2}(v_\rmn{out}^2+\sigma^2)\,,
\label{eq:power}
\end{equation}
where $v_\rmn{out}$ is the velocity of the outflowing gas and $\sigma$ is the velocity dispersion. From Fig. \ref{fig:cinematica} we have $\sigma \approx 170$
\,km\,s$^{-1}$ and adopting $v_\rmn{out}=340$\,km\,s$^{-1}$ and $M_\rmn{out}=0.3-1.1$\,M$_\odot$\,yr$^{-1}$, as discussed above, we obtain 
$\dot{E}_\rmn{out}=1.4-5.0 \times 10^{40}$erg\,s$^{-1}$, which is $\approx 0.03\%-0.1\% \times L_\rmn{bol}$, so that between 0.03\% and 0.1\% of the mass
accretion rate is transformed in kinetic power in the outflow. These values are one order of magnitude smaller than the AGN feedback derived by 
\cite{dimatteo05} in simulations to account for the co-evolution of black holes and galaxies and approximately at the same order of those estimated by 
\cite{barbosa09} and \cite{sb10} using similar IFU data. 

\begin{table}
\centering
\begin{minipage}{0.45\textwidth}
\caption{Mean stellar ages. Column 1: identification of spectra. Column 2: the spectral range, as in Table \ref{tab:population}. Column 3: aperture radius 
         in arcsec. Column 4 and 5: mean stellar age weighted by the light and mass fraction, respectively. Column 6: Emission line ratio \oiii/H$\beta$ 
         for corresponding spectrum.}
\label{tab:avgage}
\begin{tabular}{@{}cccccc@{}}
\hline\hline
Pos. & $\Delta\lambda$  &   ap.    & $<$log\,$t_*>_\rmn{L}$     & $<$log\,$t_*>_\rmn{M}$ & \oiii/H$\beta$ \\
  (1)    &        (2)   &   (3)    &           (4)              &           (5)          & (6)            \\
         &              & (arcsec) &           (yr)             &           (yr)         &                \\

\hline\hline                          
SDSS     &   $w$            &  1.5     &   8.46   &    9.94 & 8.46  \\ 
IFU 1    &   $m$            &  1.5     &   8.73   &    9.84 & 11.15 \\ 
\hline
IFU 1    &   $m$            &  0.4     &   9.48   &    9.96 & 17.9 \\
IFU 2    &   $n$            &  0.4     &   8.23   &   10.03 & 8.1  \\
IFU 3    &   $n$            &  0.4     &   7.94   &   10.03 & 7.2  \\
IFU 4    &   $n$            &  0.4     &   7.48   &    9.99 & 4.6  \\
IFU 5    &   $n$            &  0.4     &   8.22   &    9.98 & 3.4  \\
IFU 6    &   $n$            &  0.4     &   8.93   &    8.95 & 2.6  \\
IFU 7    &   $n$            &  0.4     &   6.00   &    6.00 & 0.2  \\
\hline\hline                          

\end{tabular}                
\end{minipage}               
\end{table}                 

\subsection{Stellar Population}
\label{sec:diss.stelpop}

The stellar population synthesis results are affected by the restrict wavelength range of our data, which excludes the near-UV region 
(below 4200\,\AA) -- the most sensitive to the intermediate age population. This limitation may have affected the absolute value of the contribution of this age 
component. In order to compensate for this, we have restricted the population vector to only 3 stellar components (young, intermediate age and old). The relative 
contribution of these different age components at the different locations seems to be robust, supported, for example, by the dominance of the young stellar population 
component in region 7, for which the emission-line spectrum supports ionization by stars.

From the synthesis, we conclude that within the inner 0.3\,kpc the old population dominates the flux at 4700\,\AA\ ($\sim$\,60\%), but that there is also some 
contribution of the intermediate age one ($\sim$\,30\%). The gaseous kinematics has shown that the AGN feedback is concentrated approximately within the inner 0.3\,kpc, 
coinciding with the region where the stellar population synthesis has revealed predominance of old population, but with some contribution of intermediate age population 
and FC. Beyond this region, at the 0.8\,kpc ring, the young stellar population has a contribution ranging from $\approx$\,40\% to $\approx$\,60\%, reaching 100\% at 
the detached region 7, located in an spiral arm. In regions 4, 5 and 6, there is also significant contribution of the intermediate age population.

We can represent the mixture of ages of the stellar population components in a more condensed form by its mean stellar age. Following \cite{cid05} we can define the mean 
stellar age weighted by the light fraction as
\begin{equation}
< \rmn{log}\,t_* >_\rmn{L} = \sum_{j=1}^{N_*} {x_\rmn{j}}\,{\rmn{log}\,t_\rmn{j}}
\label{eq:ageL}
\end{equation}
and by the stellar mass fraction as
\begin{equation}
< \rmn{log}\,t_* >_\rmn{M} = \sum_{j=1}^{N_*} {m_\rmn{j}}\,{\rmn{log}\,t_\rmn{j}}\,.
\label{eq:ageM}
\end{equation}
While the light-weighted mean stellar age is biased towards younger ages, as we can see in Table \ref{tab:avgage}, the mass-weighted stellar age 
is more representative of the older population (which has a larger mass-to-light ratio). According to \cite{cid05}, although the latter is more physically representative, 
it has less relation with the observed spectrum. These authors argue that, in practice,  $< \rmn{log}\,t_* >_\rmn{L}$ is the most useful of the two indices, due to the 
large M/L ratio of old stars. In Table \ref{tab:avgage} we list these two mean ages for the integrated spectrum and for the regions $1-7$. $< \rmn{log}\,t_* >_\rmn{L}$ 
is indeed characteristic of a post-starburst population at most locations, where mean ages are smaller than 1\,Gyr, except at the nucleus, where the mean age is larger 
than 3\,Gyr.

\section{Summary and Conclusions}
\label{sec:conclusions}

We have used integral field optical spectroscopy in order to map the stellar population and the gas kinematics within the inner 1.5\,kpc (radius) around the 
Post-Starburst Quasar J0212-0903 at spatial resolution of $\approx$\,0.5\,kpc (0.6 arcsec) and velocity resolution of $\approx$\,120\,km\,s$^{-1}$. This is the 
first two-dimensional study of both the stellar population and gas kinematics of a Post-Starburst Quasar and our main conclusions are:

\begin{enumerate}

\item The stellar population is dominated by old stars within 0.3\,kpc (radius) from the nucleus, while in the circumnuclear region both intermediate 
age (100\,Myrs $\le t \le$ 2.5G\,Gyrs) and young stars  ($t<100$\,Myrs) dominate the optical flux. Our results support a location of most of the post-starburst population 
in a ring with radius of $\approx$\,0.8\,kpc, where active star formation is also occurring;

\item Extended emission up to 1.5\,kpc from the nucleus is observed in both \oiii\ and H$\beta$. The \oiii\ flux distribution is more centrally concentrated, while 
the H$\beta$ flux distribution is more extended approximately along the line of the nodes and is also observed in a detached patch which seems to belong to a spiral 
arm with  active star formation;

\item The emission-line ratios are typical of Seyfert excitation within the inner 0.3\,kpc, where an outflow is observed, while beyond this region the line ratios 
are typical of  LINERs, which we attribute to a combination of diluted nuclear radiation and ionization by young stars. A Principal Component Analysis (PCA) supports 
the conclusion that recent star formation dominates the spectrum in the circumnuclear region;

\item The mass of ionized gas in the inner 1.25\,kpc (radius) is $8.34\pm0.92 \times 10^6\,\rmn{M}_\odot$;

\item The gas kinematics can be reproduced by a combination of (1) rotation in the plane of the galaxy whose line of the nodes runs approximately  North-South with 
amplitude $\le200$\,km\,s$^{-1}$ (supported also by the PCA) and (2) an outflow observed within the inner 0.3\,kpc, with blueshifts reaching up to $-670$\,km\,s$^{-1}$; 

\item  The mass outflow rate is in the range of 0.3-1.1\,M$_\odot$\,yr$^{-1}$, which is $\approx\,100$ times the AGN mass accretion rate of $\approx 8.7 \times 10^{-3}
\,\rmn{M_\odot}$\,yr$^{-1}$, implying that most of the outflow originates via mass-loading in the surrounding interstellar medium of the galaxy, swept away by the AGN 
outflow.

\end{enumerate}

We conclude that our observations support both the evolutionary and quenching scenario for this galaxy, as follows. The feeding of gas to the nuclear 
region has triggered a circumnuclear starburst a few 100's\,Myr ago, extending all the way to the nucleus. The remaining gas from this inflow, combined with the mass 
loss from the newly formed stars may have then triggered the nuclear activity, producing the observed gas outflow. This outflow, observed within the inner 0.3\,kpc, 
has then quenched further star formation at this location, in agreement with the observed absence of young stars and the contribution of intermediate age stars in 
the inner 0.3\,kpc. Beyond this region affected by the outflow, star-formation seems to be still active in a 0.8\,kpc ring. The presence of a delay between the 
triggering of star formation and the nuclear activity is supported by the presence of the intermediate age stellar population at the nucleus.

\section*{Acknowledgments}

We thank an anonymous referee for the valuable suggestions which helped to improve the paper. We thank Sabrina Lyn Cales for valuable suggestions which helped to 
improve the present paper and Rajib Gangulythe who has helped with the Gemini Proposal. Based on observations obtained at the Gemini Observatory, which is operated 
by the Association of Universities for Research in Astronomy, Inc., under a cooperative agreement with the NSF on behalf of the Gemini partnership: the National 
Science Foundation (United States), the Science and Technology Facilities Council (United Kingdom), the National Research Council (Canada), CONICYT (Chile), the 
Australian Research Council (Australia), Minist\'erio da Ci\^ncia, Tecnologia e Inova\c{c}\~ao (Brazil) and Ministerio de Ciencia, Tecnolog\'ia e Innovaci\'on 
Productiva (Argentina). This research has made use of the NASA/IPAC Extragalactic Database (NED) which is operated by the Jet Propulsion Laboratory, California 
Institute of Technology, under contract with the National Aeronautics and Space Administration. The {\small STARLIGHT} project is supported by the Brazilian 
agencies CNPq, CAPES and FAPESP, and by the France-Brazil CAPES/Cofecub program. MSB wishes to thank Brazilian agencie CAPES for the visiting professor Fellowship. 
This work has been partially supported by the Brazilian institution CNPq.


\label{lastpage}

\end{document}